%% file: paper.tex
\newcommand{\CLASSINPUTbaselinestretch}{1}
\newcommand{\CLASSINPUTbottomtextmargin}{0.8in}
\documentclass[journal]{IEEEtran}
\usepackage{etoolbox}
\makeatletter
\patchcmd{\maketitle}{\@copyrightspace}{}{}{}

\makeatother

\usepackage[]{graphicx}
\usepackage{stfloats}  
\usepackage{cite}  
\usepackage{psfrag}  
\usepackage{subfigure}
\usepackage{wrapfig}
\usepackage{epsfig}
\usepackage{url}
\usepackage{amsmath, amssymb}
\usepackage{array}
\usepackage{times}
\usepackage{multicol}
\usepackage{algorithm}
\usepackage{algorithmic}
\usepackage{mathrsfs}
\usepackage{multirow}
\usepackage{txfonts}
\usepackage[normalem]{ulem}

\usepackage[utf8x]{inputenc} 
\usepackage{ucs} 
\usepackage{amsfonts} 
\usepackage{makeidx} 
\usepackage[dvipsnames,usenames]{color} 
\usepackage{array} 
\usepackage{colortbl} 
\usepackage{booktabs} 
\usepackage[justification=centering]{caption}

\definecolor{tableheading}{rgb}{0.9,0.9,0.9}
\definecolor{softblue}{rgb}{0.8,0.8,1} 
\arrayrulecolor{ForestGreen} 
\usepackage{tabu}
\usepackage{floatflt}

\usepackage{amsmath, amssymb}
\usepackage{tikz}
\usepackage{txfonts}
\usepackage{color,soul}

\newcommand*\circled[1]{\tikz[baseline=(char.base)]{
            \node[shape=circle,fill,draw,inner sep=1.5pt,text=white,] (char) {#1};}}

\begin{document}






%

\title{RxNN: A Framework for Evaluating Deep Neural Networks on Resistive Crossbars\thanks{This work was supported in part by C-BRIC, one of six centers in JUMP, a Semiconductor Research Corporation (SRC) program sponsored by DARPA.}}
\author{\IEEEauthorblockN{Shubham Jain$^{1}$, Abhronil Sengupta$^{2}$, Kaushik Roy$^{1}$, Anand Raghunathan$^{1}$} \\
       \IEEEauthorblockA{$^{1}$School of Electrical and Computer Engineering, Purdue University \\ 
	   $^{2}$School of Electrical Engineering and Computer Science, Penn State University\\
       \{jain130,kaushik,raghunathan\}@purdue.edu, sengupta@psu.edu\\[-3.5ex]}
}

\sloppy


\maketitle


\input{sections/abstract}

\begin{IEEEkeywords}
Resistive crossbar, Non-volatile memory, Crossbar modeling, Crossbar non-idealities, Deep Neural Networks, Vector-matrix multiplication, In-memory Computing, Analog Computing, Machine Learning, Artificial Intelligence
\end{IEEEkeywords}

%
%



%
%

%
%

\vspace*{-0pt}
\section{Introduction}
\label{sec:introduction}
\input{sections/Introduction}

\vspace*{-0pt}

\vspace*{-0pt}
\section{Related Work }
\label{sec:related_work}
\input{sections/related_work}
\vspace*{-0pt}
\section{Preliminaries}
\label{sec:preliminaries}
\input{sections/preliminaries}


\vspace*{-0pt}
\section{Crossbar Non-Idealities}
\label{sec:CrossbarNI}
\input{sections/CrossbarNI}
\vspace*{-0pt}

\vspace*{-0pt}
\section{Crossbar Modeling}
\label{sec:FastCrossbarModel}
\input{sections/crossbarModeling}
\vspace*{-0pt}

\vspace*{-0pt}
\section{RxNN Framework}
\label{sec:fcm_caffe}
\input{sections/fcm_caffe}

\vspace*{-0pt}


\vspace*{-0pt}
\section{Experimental Methodology}
\label{sec:exptsetup}
\input{sections/exptsetup}

\vspace*{-0pt}
\section{Results}
\label{sec:results}
\input{sections/results}

\vspace*{-0pt}
\section{Conclusion}
\label{sec:conclusion}
\input{sections/conclusion}


\vspace*{-4pt}
\scriptsize
\bibliographystyle{unsrt}
\bibliography{references_full}
\input{biographies}

\end{document}

%% file: sections/abstract.tex
\begin{abstract}



\noindent Resistive crossbars designed with non-volatile memory devices have emerged as promising building blocks for Deep Neural Network (DNN) hardware, due to their ability to compactly and efficiently realize vector-matrix multiplication (VMM), the dominant computational kernel in DNNs.
However, a key challenge with resistive crossbars is that they suffer from a range of device and circuit level non-idealities such as driver resistance, sensing resistance, sneak paths, interconnect parasitics, non-linearities in the peripheral circuits, stochastic write operations, and process variations. These non-idealities can lead to errors in vector-matrix multiplications, eventually degrading the DNN's accuracy. It is therefore critical to study the impact of crossbar non-idealities on the accuracy of large-scale DNNs (with millions of neurons and billions of synaptic connections). However, this is challenging because existing device and circuit models are too slow to use in application-level evaluations.

We present RxNN, a fast and accurate simulation framework to evaluate large-scale DNNs on resistive crossbar systems. RxNN splits and maps the computations involved in each DNN layer into crossbar operations, and evaluates them using a Fast Crossbar Model (FCM) that accurately captures the errors arising due to crossbar non-idealities while being four-to-five orders of magnitude faster than circuit simulation. FCM models a crossbar-based VMM operation using three stages - non-linear models for the input and output peripheral circuits (Digital-to-Analog and Analog-to-Digital converters), and an equivalent non-ideal conductance matrix for the core crossbar array. 
We implement RxNN by extending the Caffe machine learning framework and  use it to evaluate a suite of six large-scale DNNs 
developed for the ImageNet Challenge (ILSVRC). Our experiments reveal that resistive crossbar non-idealities can lead to significant accuracy degradations (9.6\%-32\%) for these large-scale DNNs. To the best of our knowledge, this work is the first quantitative evaluation of the accuracy of large-scale DNNs on resistive crossbar based hardware. We also demonstrate that RxNN enables fast model-in-the-loop re-training of DNNs to partially mitigate the accuracy degradation.

\end{abstract}



%% file: sections/Introduction.tex
\noindent Deep neural networks (DNNs) have transformed the field of artificial intelligence in the past decade, and are currently used in several real-world products and services for speech recognition, image analysis, natural language processing, search engines, recommendation systems, and more~\cite{fortune-dnns,wired-dnns}. 
However, the large and rapidly growing computation and storage requirements of DNNs pose severe challenges to the systems on which they are deployed.

Resistive crossbars have garnered significant interest in realizing DNNs due to their ability to perform the underlying computational kernel, {\em viz.} vector-matrix multiplication (VMM), efficiently. They may be designed using a range of emerging devices, including Resistive RAM (ReRAM), Phase Change Memory (PCM), and Spintronics~\cite{MemristorAsSynapse,yongpangProspects,sengupta_synapseMain,dfan_dtcs_res}. These devices have several desirable characteristics such as high density, non-volatility, and low voltage operation, enabling highly compact and energy-efficient DNN implementations. Consequently, several research efforts have explored resistive crossbar based hardware at various levels of design abstraction~\cite {spindle,prime,reno,ISAAC,BSB_HelenLi,harmonica,time,dpeHP_DAC16,closedLoop_Hli,jiangLi_NNtraining,kataeva2015,mitigatingNIsynapticDevice,tayfunMLP,tayfunCNN,chakraborty2018technology,trainingItself_YChen,noise_elimination_Training_YChen,irdrop_ychen,groupScissor,rescuingDefects,mnsim,NeuroSIM,rram_exploration,AutoNCS,Liu_EDA,NEUTRAMS,ibm_RPU,jiangLi_sneakPaths,yongpangProspects,xMANN,sJainIBMJournal,cxdnn}.


A key challenge with resistive crossbars is that the performed operation is only an approximation of the desired vector-matrix multiplication. For example, consider the process where digital inputs are first converted to voltages and applied to the rows of the crossbar (which is programmed with the weights as conductances), and the resulting column currents are digitized to generate the digital outputs. Various device and circuit level non-idealities,~\emph{viz.}, driver resistance, sensing resistance, sneak paths, interconnect parasitics, ADC and DAC non-linearity, stochastic write operations, and process variations may lead to errors in the computed vector-matrix multiplications. These errors can degrade the overall accuracy of a DNN realized on a resistive crossbar system. Although DNNs are resilient to some imprecision in their computations~\cite{AxNN,PrecisionLimited_IBM,8bitImagenetWithoutMultiplier}, this resilience is not unlimited. Therefore, it is necessary to evaluate the impact of resistive crossbars non-idealities at the application level in order to establish their feasibility as the building blocks of DNN hardware.

Most previous efforts on resistive crossbar based DNN implementations either do not consider non-idealities or model non-idealities in a very limited manner ({\em e.g.}, as limited precision). Moreover, they focus their analysis on very simple networks and datasets ({\em e.g.}, MNIST). Thus, they leave open the question of how non-idealities impact the accuracy of \emph{large-scale neural networks} and more complex tasks. State-of-the-art DNNs often contain tens to hundreds of layers, millions of neurons and billions of synaptic connections. Evaluating such networks requires a fast and scalable, yet accurate, model for resistive crossbars that can be integrated into state-of-the-art DNN software frameworks. Unfortunately, such a framework is currently unavailable. Device and circuit simulation (SPICE) models of resistive crossbars are accurate but extremely slow and infeasible for large-scale network evaluation. On the other hand, architectural models of resistive crossbars~\cite{mnsim,rram_exploration} target design space exploration and use highly simplified error models that are reasonable for their context, but inadequate for evaluating application-level accuracy of DNNs. For example, these models do not consider error dependence on the crossbar inputs, programmed conductances, and the crossbar column performing the computation.

We propose RxNN, a fast and accurate simulation framework to enable functional evaluation of large-scale DNNs on resistive crossbars. RxNN splits and maps the DNN's computations into crossbar operations, and evaluates them using a Fast Crossbar Model (FCM) that accurately captures the errors arising due to crossbar non-idealities while being four-to-five orders of magnitude faster than circuit simulation. FCM models a crossbar-based VMM operation using three stages - non-linear models for the input and output peripheral circuits (Digital-to-Analog and Analog-to-Digital converters), and an equivalent non-ideal conductance matrix for the core crossbar array. The non-ideal conductance matrix for each crossbar array is derived by pre-solving the voltage-current equations (Kirchoff's loop law and Ohm's law) for the array, requiring only a vector-matrix multiplication to subsequently transform the input voltages to output currents.

We realize FCM using the well-known BLAS (Basic Linear Algebra Subprograms) interface and develop RxNN based on the Caffe~\cite{caffe} deep learning framework. We use RxNN to evaluate six  large-scale DNNs for classifying the ImageNet~\cite{imagenet} dataset and three simpler networks for classifying the CIFAR-10 and MNIST datasets. Our evaluation suggests that crossbar non-idealities impact accuracy much more significantly as we move from the smaller networks (LeNet and ConvNet) and simpler datasets to the larger networks ({\em e.g.}, ResNet) and ImageNet,
motivating the need for further research in cross-layer mitigation and compensation techniques.

In summary, the key contributions of this work are:
\begin{itemize}
\item We study the cumulative effect of resistive crossbar non-idealities by characterizing the resulting errors in the realized vector-matrix multiplication operations. We find that the errors show significant data and hardware-instance dependence that should be considered for accurately modeling resistive crossbars. 
\item We propose FCM, a fast and accurate functional crossbar model to capture the effects of crossbar non-idealities.   
\item We develop RxNN, a software framework that can evaluate large-scale DNNs on resistive crossbar systems and help re-train to compensate for the effects of non-idealities. 
\item We evaluate the application-level accuracy of 6 state-of-the-art DNNs, \emph{viz.} ResNet-50, VGG-16, GoogleNet, AlexNet, OverFeat, and NiN on a resistive crossbar based system. Our evaluation reveals that the degradation in accuracy due to non-idealities can be significant (9.6\%-32\%) for large-scale DNNs. This degradation can be partially alleviated by re-training, but calls for further research in compensation techniques.
\end{itemize}

The rest of the paper is organized as follows. Section~\ref{sec:related_work} overviews the prior efforts related to this work. Section~\ref{sec:preliminaries} provides the necessary background on resistive crossbars. Section~\ref{sec:CrossbarNI} discusses crossbar non-idealities and demonstrates their impact on vector-matrix multiplications realized using crossbars. Section~\ref{sec:FastCrossbarModel} describes the proposed FCM models. Section~\ref{sec:fcm_caffe} presents the RxNN software framework. Section~\ref{sec:exptsetup} details the experimental methodology. We present experimental results in section~\ref{sec:results} and conclude the paper in Section~\ref{sec:conclusion}.

%% file: sections/related_work.tex
Resistive crossbars have attracted significant interest in recent years due to their ability to efficiently realize vector-matrix multiplications, the key computational kernel in DNNs~\cite{neuromorphicHardware_Survey,LuSynapse,wongRRAM,SharadNeuron}. Prior efforts towards realizing DNNs on resistive crossbar systems can be broadly classified into specialized hardware accelerator designs~\cite{spindle,prime,reno,harmonica,time,BSB_HelenLi}, non-ideality mitigation schemes~\cite{jiangLi_NNtraining,dpeHP_DAC16,kataeva2015,closedLoop_Hli,mitigatingNIsynapticDevice,groupScissor,rescuingDefects,noise_elimination_Training_YChen,trainingItself_YChen,irdrop_ychen}, and design tools for resistive crossbar systems~\cite{mnsim,NeuroSIM,rram_exploration,AutoNCS}.  

\vspace*{+4pt}
{\bf\noindent Specialized hardware accelerators.} Resistive crossbar based specialized hardware systems have been proposed for accelerating DNN inference~\cite{spindle,prime,reno,BSB_HelenLi,ISAAC} and training~\cite{harmonica,time}. These efforts focus on the evaluation of the proposed architecture using performance, energy, and area as their metrics, and either do not consider non-idealities or model only the quantized nature of crossbar-based VMM operations.

\vspace*{+4pt}
{\bf\noindent Non-ideality mitigation schemes.} Prior efforts have also proposed methods to mitigate the impact of crossbar non-idealities. These include (i) (re-)training~\cite{kataeva2015,mitigatingNIsynapticDevice,jiangLi_NNtraining,tayfunMLP,tayfunCNN,chakraborty2018technology}, (ii) optimized weight to conductance conversion~\cite{dpeHP_DAC16}, (iii) rank clipping to reduce the effects of non-idealities by lowering crossbar dimensions~\cite{groupScissor}, (iv) schemes to alleviate the effect of hard failures~\cite{rescuingDefects}, and (vi) hardware solutions to address low-voltage induced drift~\cite{closedLoop_Hli}, programming errors~\cite{noise_elimination_Training_YChen}, and IR drop~\cite{irdrop_ychen}. The focus of all these efforts is to evaluate and mitigate errors due to crossbar non-idealities. However, they are restricted to simple networks and small datasets. This limitation is in large part due to the lack of a scalable simulation framework. Further, many of these efforts also lack a detailed crossbar model and consider only a subset of crossbar non-idealities.

RxNN complements the above efforts on hardware accelerators and non-ideality mitigation schemes. Moreover, it overcomes their limitations (such as considering only a limited set of non-idealities), and can accurately model crossbar-based systems while maintaining very high simulation speed (several orders-of-magnitude faster than SPICE). 


\vspace*{+4pt}
{\bf\noindent Design tools.} To aid design space exploration, prior efforts~\cite{mnsim,NeuroSIM,rram_exploration,AutoNCS} have proposed circuit-level macro models to evaluate crossbar systems. These efforts include (i) MNSIM~\cite{mnsim}, a simulation platform to evaluate inference accelerators designed using resistive crossbars, (ii) NeuroSim~\cite{NeuroSIM}, a framework to evaluate crossbars systems designed for on-chip training, (iii) a technology exploration tool for resistive crossbars~\cite{rram_exploration}, and (iv) AutoNCS~\cite{AutoNCS}, a tool to optimize the utilization and efficiency of a resistive crossbar system. While the primary focus of these tools has been performance, energy, and area evaluation of resistive crossbar systems, they also include simplistic accuracy/error models that are reasonable for design space exploration, but inadequate for evaluating application-level accuracy of large-scale DNNs. RxNN complements these efforts by focusing on accurately and efficiently modeling crossbar non-idealities in the context of large-scale DNNs.

%% file: sections/preliminaries.tex
In this section, we provide a brief background on resistive crossbars and how they evaluate vector-matrix multiplication.

\begin{figure}[htb]
  \centering
  \vspace*{-10pt}
  \includegraphics[width=\columnwidth]{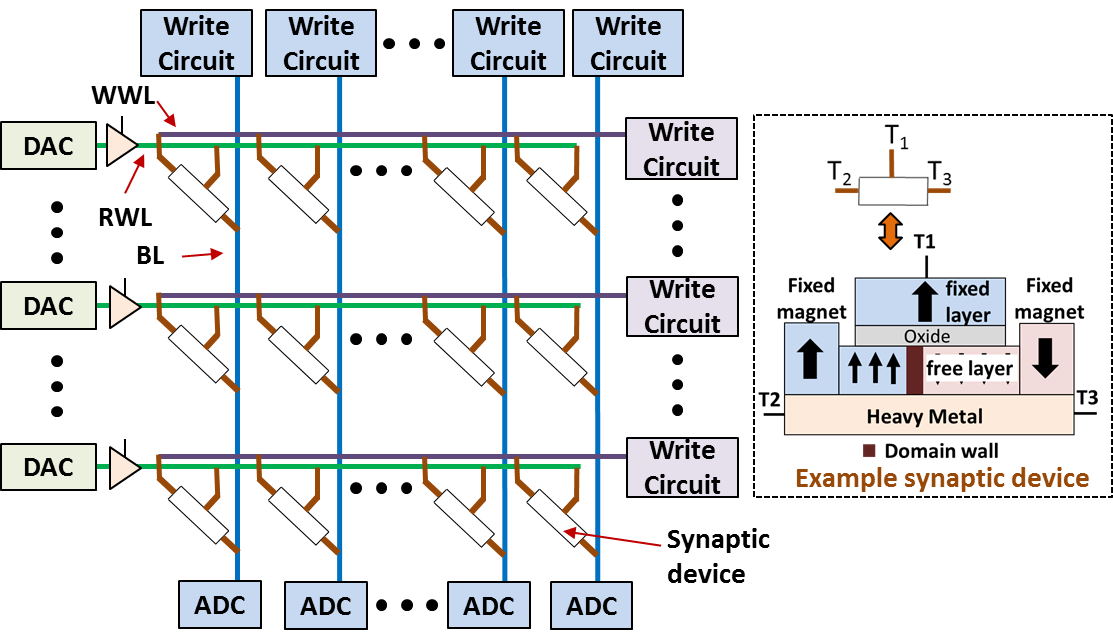}
  \vspace*{-8pt}
  \caption{Resistive Crossbar array}
  \label{fig:preliminaries}
  \vspace*{-8pt}
\end{figure}

Figure~\ref{fig:preliminaries} presents a typical resistive crossbar array design for realizing vector-matrix multiplications. It consists of a 2D array of synaptic devices, Digital-to-Analog converters (DACs), and Analog-to-Digital converters (ADCs), and write peripheral circuitry. It supports two main operations: (i) Programming, \emph{i.e.}, a write operation performed sequentially on a subset of synaptic devices, and (ii) Evaluation, {\em i.e.}, the vector-matrix multiply operation. The synaptic element at the intersection of each row and column is programmed by enabling the corresponding write circuits along the Write Wordline (WWL) and the bitline (BL), to apply the necessary current and set it to the desired conductance~\footnote{although not shown in the figure, in general, an access transistor or selector device may be required in addition to the synaptic device for write operations.}. A vector-matrix multiplication is performed by using DACs to convert the digital inputs into voltages on the RWLs, and sensing the resulting current flowing through the BLs using ADCs.  

Synaptic devices are programmable resistors that are commonly realized using emerging non-volatile memory technologies such as PCM, ReRAM, and Spintronics~\cite{MemristorAsSynapse,yongpangProspects,sengupta_synapseMain,dfan_dtcs_res}. Figure~\ref{fig:preliminaries} illustrates an example of a spintronic synaptic device~\cite{sengupta_synapseMain} consisting of a Magnetic Tunneling Junction (MTJ) and an underlying Heavy Metal (HM) layer. The 3-terminal device is programmed though the HM layer and sensed through the MTJ. The position of the domain wall determines the conductance of the MTJ that lies between $G_{MIN}$ (when the domain wall is to the far right) and $G_{MAX}$ (when the domain wall is to the far left). The number of unique locations at which the domain wall can reside determines the precision of the device. Although we consider this spintronic synaptic device in our explanations, the RxNN framework is applicable to the wide range of resistive devices used to design crossbars.

Equation~\ref{eq:equation_ideal} specifies the ideal vector-matrix multiply operation for an MxN dimensional crossbar. $Vin_{ideal}$ is a 1xM vector consisting of the input voltages, $G$ is an MxN matrix comprising of the synaptic conductances, and $Iout_{ideal}$ is a 1xN vector containing the output currents. 

\vspace*{-0pt}
\begin{equation}
\centering
\begin{aligned}
   \begin{split}
 	& \textbf{Iout}_\textbf{ideal} = \textbf{Vin}_\textbf{ideal} *\textbf{G}_\textbf{ideal}
	\end{split}
	\end{aligned}
	\label{eq:equation_ideal}
\vspace*{-10pt}
\end{equation}


\begin{figure*}[htb]
  \centering
  \vspace*{-0pt}
  \includegraphics[width=\textwidth]{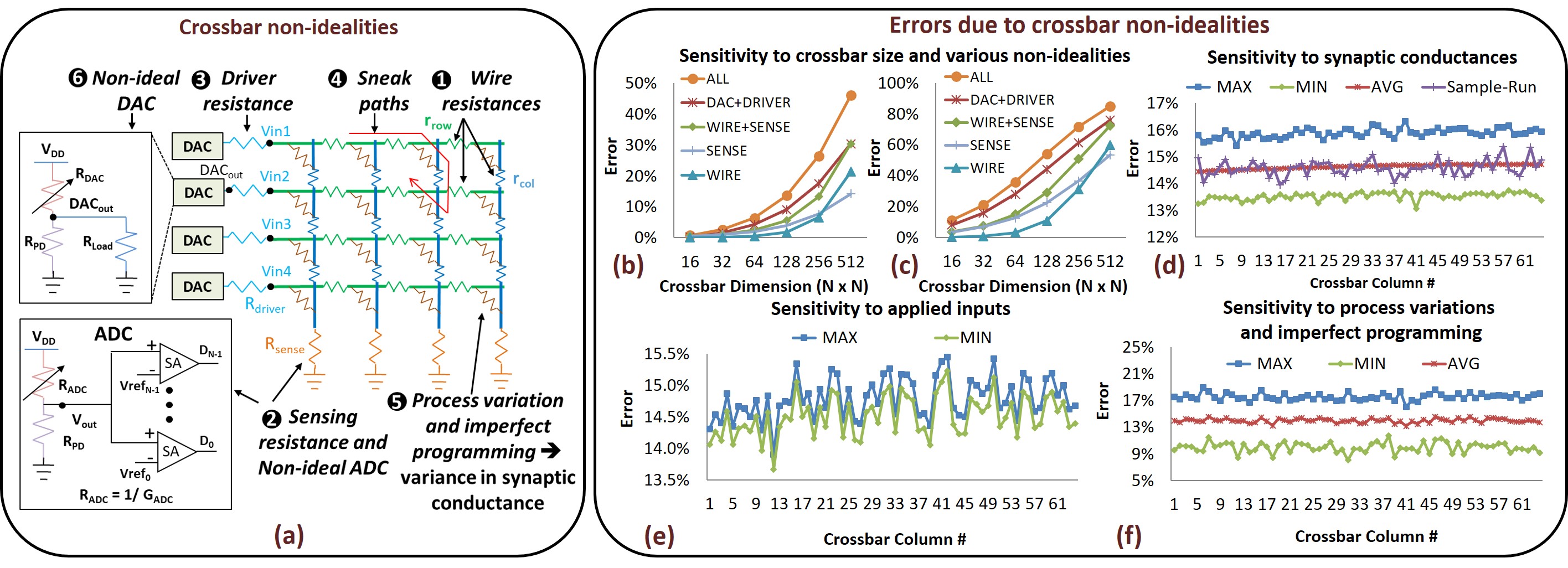}
  \vspace*{-10pt}
  \caption{ Crossbar non-idealities: (a) Resistive equivalent circuit for a crossbar, (b)-(c) Sensitivity to crossbar dimensions with all synaptic conductances programmed to $G_{MIN}$ and $G_{MAX}$, respectively, (d) Sensitivity to synaptic conductances, (e) Sensitivity to applied inputs, and (f) Sensitivity to process variation and imperfect programming}
  \label{fig:NI_evaluation}
  \vspace*{-15pt}
\end{figure*}

%% file: sections/CrossbarNI.tex
\noindent In this section, we analyze non-idealities in resistive crossbars and examine their impact on vector-matrix multiplications.

\subsection{Crossbar Non-idealities}
\label{subsec:NI_overview}

Figure~\ref{fig:NI_evaluation}(a) presents an equivalent circuit for the crossbar array and the peripherals (DACs and ADCs) that includes various non-idealities. The key sources of non-idealities are ---~\circled{1} wire resistances of the crossbar interconnects,~\circled{2} sensing resistances of the circuits that sense the output currents,~\circled{3} driver resistances of the circuits that drive the crossbar rows,~\circled{4} sneak paths,~\circled{5} variance in synaptic conductance due to process variations and imperfect programming, and ~\circled{6} non-ideal DACs. While we consider all these non-idealities in subsequent sections, we select the non-idealities due to DACs and sneak paths for a more detailed treatment below, in order to illustrate the complexity of error modeling.

\begin{figure}[htb] 
  \centering
  \vspace*{-0pt}
  \includegraphics[width=\columnwidth]{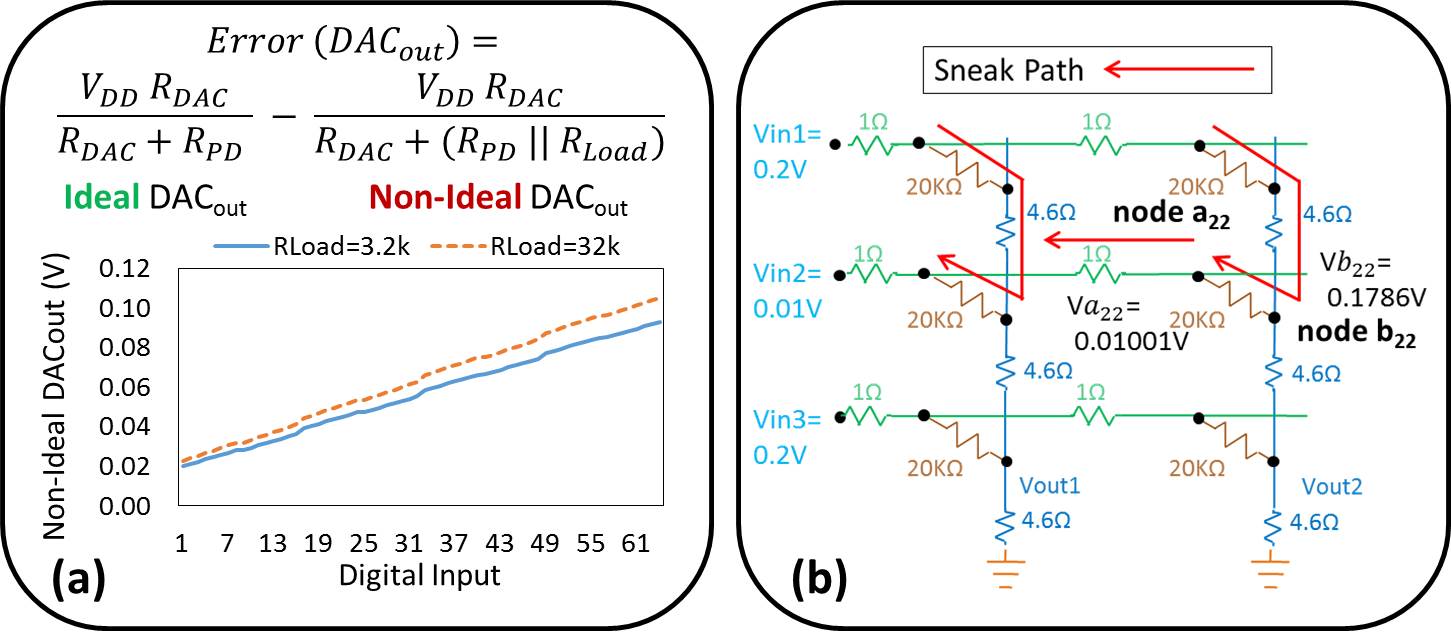}
  \vspace*{-10pt}
  \caption{Example of non-idealities in resistive crossbar}
  \label{fig:NIExample}
  \vspace*{-10pt}
\end{figure}

\vspace*{+4pt}
{\bf\noindent Non-ideal DAC.} Figure~\ref{fig:NI_evaluation}(a) shows the equivalent  circuit for a DAC that is represented using a resistive divider circuit with an input determined resistance ($R_{DAC}$) and a fixed resistance ($R_{PD}$). An applied digital input determines the value of $R_{DAC}$ and subsequently decides the DAC's output voltage ($DAC_{out}$). Note that, $DAC_{out}$ also depends on the effective resistive load ($R_{Load}$), leading to deviations from the ideal value. $R_{Load}$ is a function of the synaptic conductances within the crossbar array and therefore varies with the crossbar state (the values of all synaptic conductances). The equation in Figure~\ref{fig:NIExample}(a) shows the error incurred due to DAC non-idealities which is a function of both applied inputs ($R_{DAC}$) and synaptic conductances ($R_{Load}$). Figure~\ref{fig:NIExample}(a) also plots the non-ideal DAC output for two load resistances, {\em viz.} 3.2k$\Omega$ and 32k$\Omega$, respectively. As evident, the DAC output voltage varies significantly with the load for any given input. 


\vspace*{+4pt}
{\bf\noindent Sneak paths during vector-matrix multiplication.} Ideally, currents in resistive crossbars would be expected to flow from left to right along the rows and from top to bottom through the columns. However, due to the non-idealities described above (specifically, wire resistances), internal node voltages within the crossbar may vary, resulting in additional current paths, which we refer to as sneak paths. Figure~\ref{fig:NIExample}(b) illustrates sneaks paths during vector-matrix multiplications for a 3x2 crossbar array. We consider a crossbar state with all synaptic devices programmed to 20K$\Omega$, and the applied input voltages at the rows are 0.2V, 0.01V and 0.2V, respectively. For this crossbar state, we observe that the direction of current between nodes $a_{22}$ and $b_{22}$ is flipped, \emph{i.e.}, the current flows from $b_{22}$ towards the input (Vin2), instead of the expected direction. Sneak paths are a function of both the crossbar state and the applied inputs, and therefore further contribute to the overall dynamism in errors due to non-idealities.

\subsection{Errors due to Non-Idealities}
\label{subsec:NI_evaluation}

Next, we study the impact of non-idealities on the computational accuracy of the vector-matrix multiplication realized using resistive crossbars. To this end, we compare the outputs of vector-matrix multiplications obtained from HSPICE simulations of non-ideal crossbar arrays with the ideal computations (Equation~\ref{eq:equation_ideal}) and analyze the sensitivity of the errors to various parameters. 

\vspace*{+4pt}
{\bf\noindent Sensitivity to crossbar size.}  We first examine how the errors incurred due to individual non-idealities (WIRE, SENSE), combinations of non-idealities (DAC+DRIVER, WIRE+SENSE), and the cumulative effect of all non-idealities (ALL) vary with the crossbar dimension. Figures~\ref{fig:NI_evaluation}(b) and~\ref{fig:NI_evaluation}(c) show the errors incurred during the vector-matrix multiplication realized using crossbars, with all synaptic conductances programmed to $G_{MIN}$ and $G_{MAX}$, respectively. In both graphs, the Y-axis represents the error in the last ($N^{th}$) column of an NxN crossbar, and the X-axis represents the crossbar dimension (N). In both cases, we observe that the errors due to all non-idealities (ALL), individual non-idealities, and subsets of non-idealities, increase with the crossbar dimension. This is expected because: (i) the overall wire resistances increase with crossbar array size, (ii) the sensing resistance contribution to the overall bitline resistance increases, and (iii) the DAC non-ideality increases due to a decrease in the effective load resistance~\footnote{Higher crossbar dimensions have more columns leading to increase in parallel paths, consequently lowering the effective load resistance}. Further, we also observe that for smaller crossbars, the non-ideality due to DAC is predominant, whereas, for larger crossbars, the wire and sensing resistance effect becomes equally significant. 

\vspace*{+4pt}
{\bf\noindent Sensitivity to crossbar state.} Next, we characterize errors' dependence on the crossbar state, \emph{i.e.}, the conductances of all synaptic devices. To this end, we fix the inputs to a 64x64 crossbar array and vary the conductances of the synaptic devices to obtain different crossbar states. Figure~\ref{fig:NI_evaluation}(d) shows the maximum (MAX), minimum (MIN), and average (AVG) errors across columns of the crossbar over 1000 random crossbar states. We observe that the errors show significant variation across these states. In Figure~\ref{fig:NI_evaluation}(d), we also plot the errors for a sample crossbar state (Sample-Run) to demonstrate the variation of errors across crossbar columns. Note that this pattern is quite different from the patterns observed for MAX, MIN, and AVG errors. 

\vspace*{+4pt}
{\bf\noindent Sensitivity to crossbar inputs.} To analyze the errors' dependence on the applied inputs, we fixed the conductances of all synaptic devices and varied the inputs. Figure~\ref{fig:NI_evaluation}(e) shows the variations in errors across inputs. We observe that the variance across inputs (MAX and MIN) for a particular column is noticeable, but small in comparison to the variance across crossbar states. 
 
\vspace*{+4pt}
{\bf\noindent Sensitivity to crossbar columns.} Figures~\ref{fig:NI_evaluation}(d-e) depicts how errors vary across crossbar columns. While there is a slight trend of increase in error as we go from the first to the last column, it is not always the last column that incurs the maximum error. Rather any column can incur the maximum error depending on the crossbar state and the applied inputs.   

\vspace*{+4pt}
{\bf\noindent Sensitivity to process variation and imperfect programming.} Finally, we also evaluate the impact of variations by performing Monte-Carlo simulation on a sample set of 10,000 crossbar states obtained by considering variations in synaptic conductances ($\sigma$/$\mu$ = 10\%)~\cite{synapse_variation}. Figure~\ref{fig:NI_evaluation}(f) shows the maximum, minimum, and average error observed on a 64x64 crossbar array across these samples. The variations in synaptic conductances can occur due to two prominent reasons: (i) Process variations and (ii) Imperfect programming, \emph{i.e.}, errors during write operations. 

In summary, the non-idealities in resistive crossbars can have a significant impact on the computations that they perform. Furthermore, the errors due to non-idealities are highly dependent on various factors, including the conductances, applied inputs, crossbar column and process variations. Thus, a crossbar model should consider these factors in order to accurately capture the impact of non-idealities on application-level accuracy. 

%% file: sections/crossbarModeling.tex
In this section, we present a Fast Crossbar Model (FCM) that accurately captures the impact of resistive crossbars non-idealities on the performed vector-matrix multiplications.

\subsection{FCM Overview}
\label{subsec:fcmoverview}

\begin{figure}[htb]
  \centering
  \vspace*{-0pt}
  \includegraphics[width=\columnwidth]{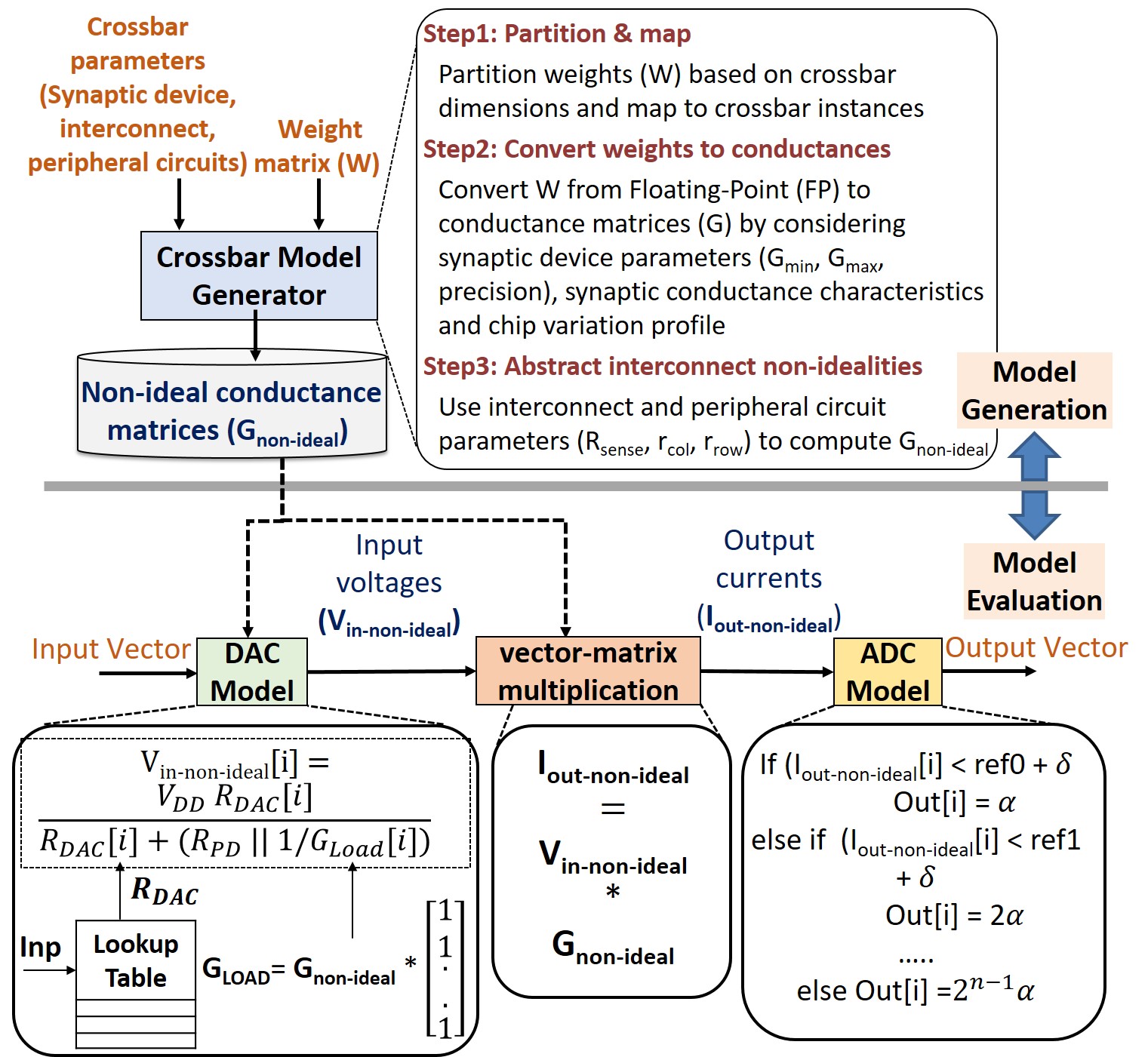}
  \vspace*{-10pt}
  \caption{FCM: Overview}
  \label{fig:fcmoverview}
  \vspace*{-10pt}
\end{figure}

Figure~\ref{fig:fcmoverview} overviews the proposed fast crossbar model that consists of two phases: (i) Model generation and (ii) Model evaluation. Model generation is a one-time step for each DNN, whereas model evaluation is invoked to evaluate each inference operation using the DNN. The key idea FCM is to first abstract non-idealities in the core crossbar array by transforming a weight matrix (W) into a non-ideal conductance matrix ($G_{non-ideal}$). Subsequently, using the generated $G_{non-ideal}$ matrix and non-linear peripheral (ADC and DAC) models, FCM emulates the non-ideal vector matrix multiplications on resistive crossbars. We discuss these phases of FCM in detail below.

\begin{figure}[htb]
  \centering
  \vspace*{-10pt}
  \includegraphics[width=\columnwidth]{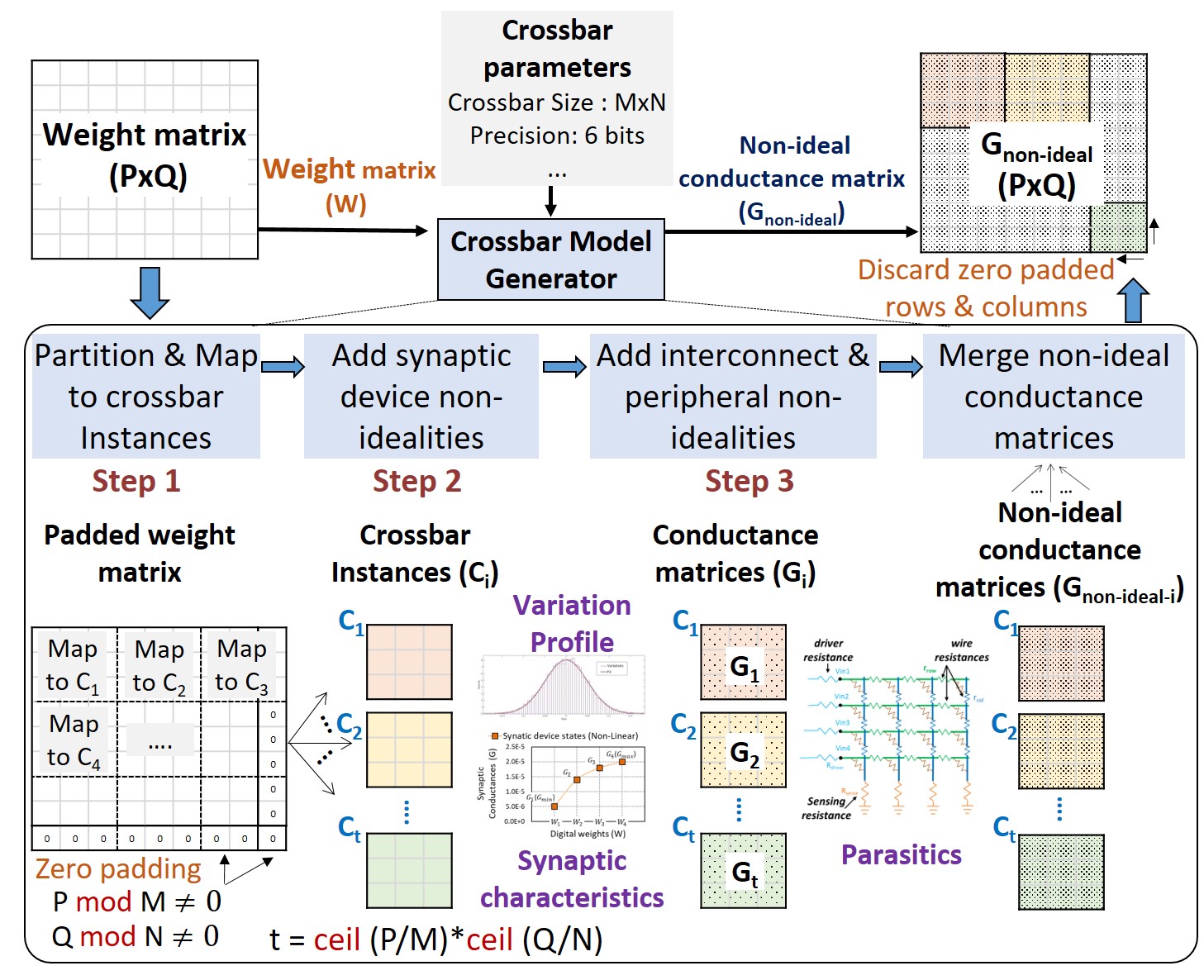}
  \vspace*{-10pt}
  \caption{Crossbar model generator: Details}
  \label{fig:mapping}
  \vspace*{-10pt}
\end{figure}

\begin{figure}[htb]
  \centering
  \vspace*{-5pt}
  \includegraphics[width=\columnwidth]{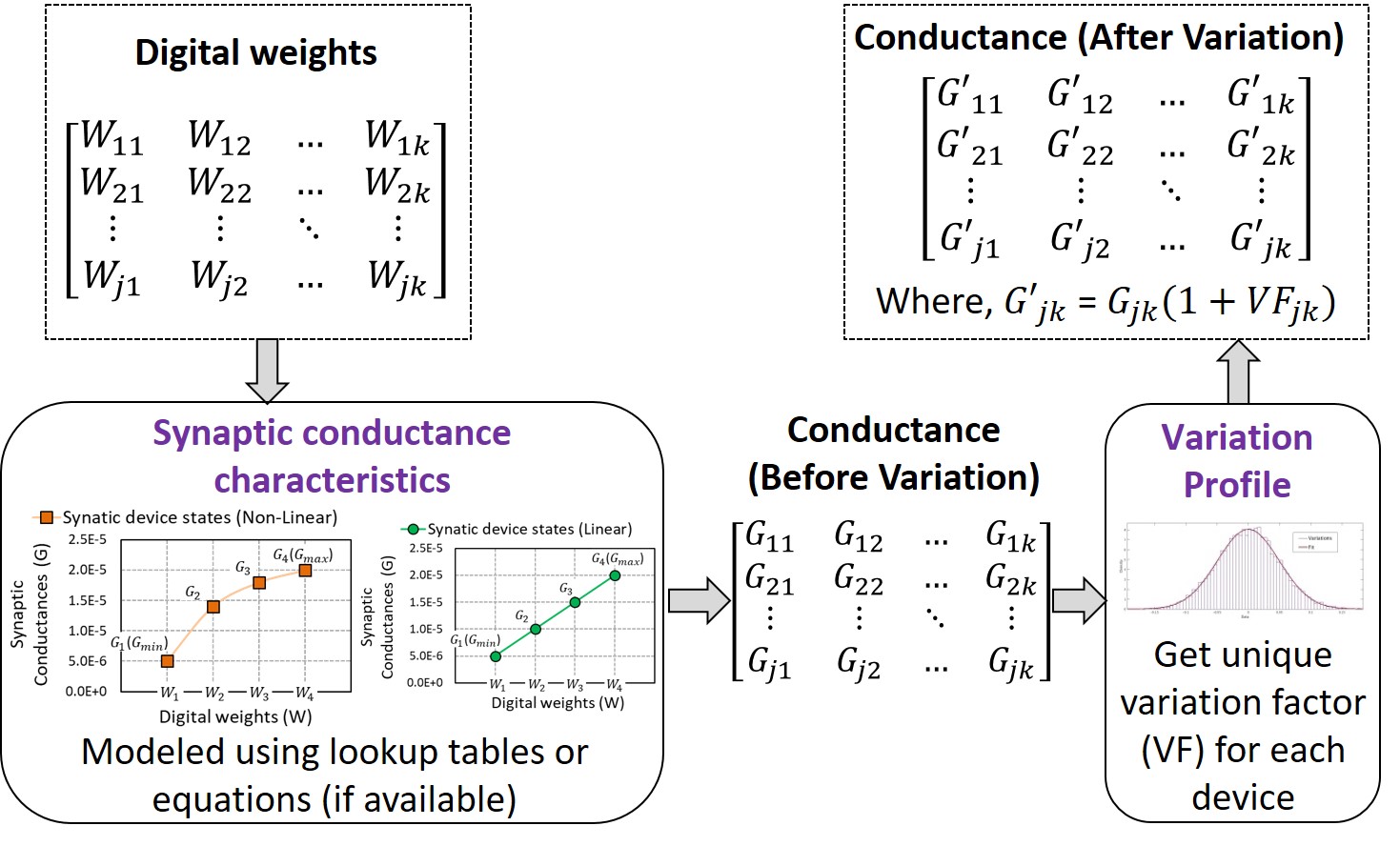}
  \vspace*{-10pt}
  \caption{Abstracting synaptic device non-idealities}
  \label{fig:deviceNI}
  \vspace*{-10pt}
\end{figure}

{\bf \noindent Crossbar model generator.} FCM uses a crossbar model generator to abstract the non-idealities arising due to  process variations, the non-linear synaptic conductance characteristics, the sensing resistance, and the wire resistances. The model generator takes crossbar parameters and a weight matrix (W) as inputs and generates a non-ideal conductance matrix ($G_{non-ideal}$) as the output. Using a three-step transformation mechanism (listed in Figure~\ref{fig:fcmoverview}), it converts W to $G_{non-ideal}$ based on crossbar parameters including synaptic device characteristics ($G_{min}$, $G_{max}$, precision), interconnect ($R_{Sense}$, $r_{row}$, $r_{col}$), and circuit (crossbar size) parameters, and the chip variation profile. Figure~\ref{fig:mapping} illustrates the model generation process in greater detail using an example, where we consider the mapping of a PxQ weight matrix to crossbars of size MxN. In step 1, the model generator slices the matrix W into fragments and maps them to specific crossbar instances. The fragment size is same as the crossbar dimension and to achieve this for all fragments the corners of the matrix W are zero padded if required. Note that, we need t=$\bf{ceil}$(P/M)*$\bf{ceil}$(Q/N) crossbar instances, `P $\bf{mod}$ M' zero-padded rows, and `Q $\bf{mod}$ N' zero-padded columns. Next, in step 2 (described in Figure~\ref{fig:deviceNI}), weights are converted from digital floating-point (FP) values to conductances (G) considering device conductance characteristics, parameters ($G_{min}$, $G_{max}$, precision), and the variation profile. FCM supports synaptic devices with both linear~\cite{sengupta_synapseMain,LGao-NanoTech15} and non-linear~\cite{LuSynapse,mitigatingNIsynapticDevice} conductance characteristics, either using equations or lookup tables. We sample the variation profile to obtain a unique variation factor (VF) for each synaptic element within each crossbar instance. At the end of step 2, we obtain a conductance matrix ($G_{i}$) for each crossbar instance. Finally, in step 3, the generator abstracts interconnect non-idealities ($R_{sense}$, $r_{col}$, $r_{row}$) and transforms the conductance matrices ($G_{i}$) to the corresponding non-ideal conductance matrices ($G_{non-ideal-i}$). Subsequently, these non-ideal conductance matrices ($G_{non-ideal-i}$) are merged to obtain one $G_{non-ideal}$ matrix corresponding to the weight matrix $W$. The transformation of $G_{i}$ to $G_{non-ideal-i}$ is exact, and we provide the mathematical proof in Section~\ref{subsec:fcmDerivation}. 

\vspace*{+4pt}

{\bf\noindent Peripheral (ADC \& DAC) models.} Figure~\ref{fig:fcmoverview} details the ADC and DAC models used by FCM to incorporate ADC and DAC non-idealities. The DAC model is composed of a resistive divider circuit with a digital input (Inp) dependent resistance ($R_{DAC}$) and a fixed resistance ($R_{PD}$). The resistive divider is connected to a variable effective load conductance ($G_{Load}$) whose value is dependent on the crossbar state (synaptic conductances). FCM uses the equation shown in Figure~\ref{fig:fcmoverview} to compute the non-ideal input voltages ($V_{in-non-ideal}$). In the equation, $R_{DAC}$ is determined using the digital inputs (Inp), and $G_{Load}$ is computed using the $G_{non-ideal}$ matrix. We note that $V_{in-non-ideal}$ captures the data-dependence of the errors arising due to a non-ideal DAC, since $R_{DAC}$ and $G_{Load}$ are dependent on the applied inputs and the crossbar state, respectively. Using matrices $G_{non-ideal}$ and $V_{in-non-ideal}$, FCM computes the non-ideal vector-matrix multiplication realized in crossbars to obtain non-ideal output currents ($I_{out-non-ideal}$). The ADC model shown in Figure~\ref{fig:fcmoverview} (which can model non-linearity) is then used to convert the $I_{out-non-ideal}$ to digital outputs (Out). 

In summary, FCM abstracts both device and circuit non-idealities into crossbar models that achieve several orders-of-magnitude speed-up over SPICE, without compromising on the modeling accuracy (in our experiments, FCM models are functionally within 0.28\% of HSPICE). We note that achieving such simulation speed is not possible without some abstraction, and that FCM provides a good tradeoff between fidelity vs. simulation speed (detailed in section~\ref{subsec:fcmresult}). FCM further derives simulation speed by
realizing algebraic operations using well-optimized BLAS (Basic Linear Algebra Subprograms) libraries.


\begin{figure}[htb]
  \centering
  \vspace*{-5pt}
  \includegraphics[width=\columnwidth]{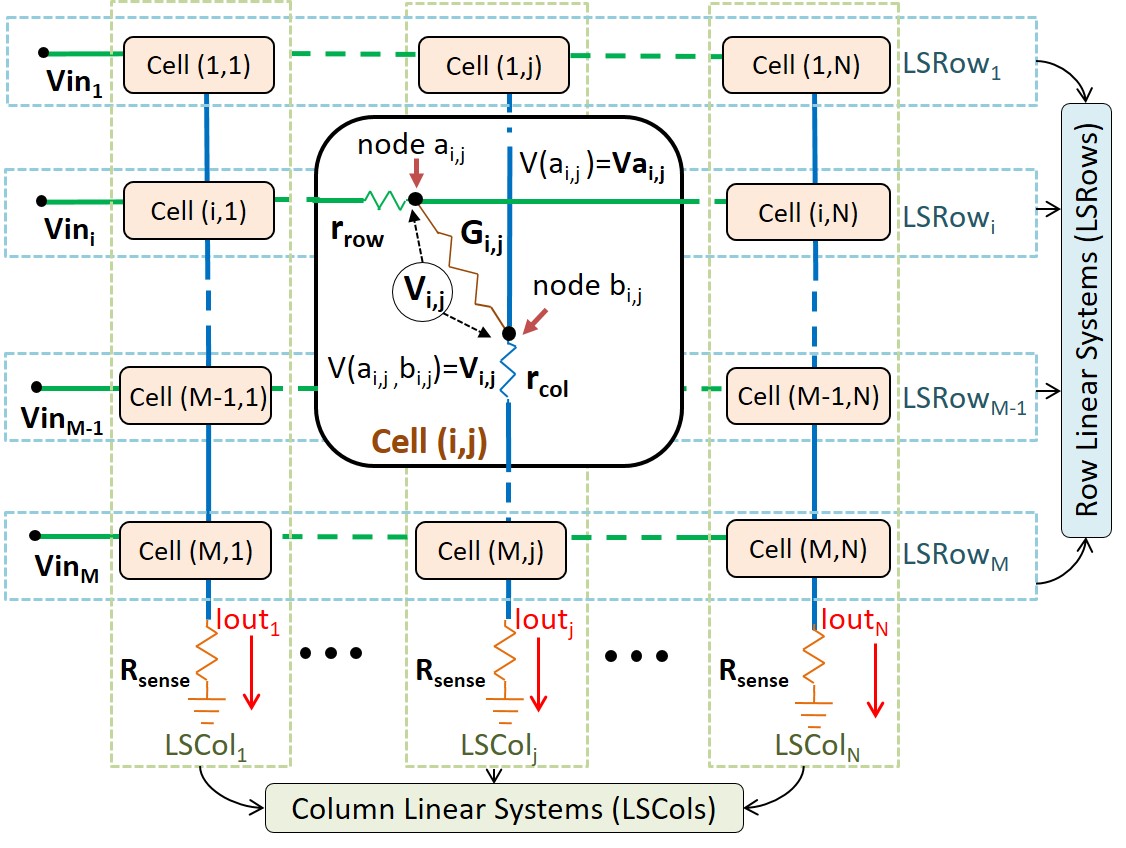}
  \vspace*{-10pt}
  \caption{ Equivalent resistance circuit of MxN crossbar array}
  \label{fig:resEq}
  \vspace*{-10pt}
\end{figure}



\subsection{Abstraction of interconnect non-idealities}
\label{subsec:fcmDerivation}

In this section, we provide the mathematical formulation for the abstraction of interconnect non-idealities (Step 3 of crossbar model generation). We recall that in this step the generator abstracts interconnect non-idealities ($R_{sense}$, $r_{col}$, $r_{row}$) and transform the conductance matrix ($G_{i}$) associated with the $i^{th}$ crossbar instance to the corresponding non-ideal conductance matrix ($G_{non-ideal-i}$). We achieve this transformation by leveraging circuit laws (Kirchhoff's loop laws and Ohm's law) and linear algebraic operations (direct sum, row switching, vector concatenation, row reduction, etc.).

We now explain the formulation using Figure~\ref{fig:resEq} that shows the equivalent resistive circuit of an MxN crossbar array. $Vin_{i}$ represents the input voltage at the $i^{th}$ row of the crossbar, $Va_{i,j}$ denotes the voltage at the node $a_{i,j}$, and $V_{i,j}$ is the voltage difference between the node $a_{i,j}$ and the node $b_{i,j}$. $G_{i,j}$ is the conductance of the synaptic device at the $i^{th}$ row and the $j^{th}$ column. $R_{sense}$, $r_{row}$, and $r_{col}$ depict the sensing and distributed wire resistances, respectively, and $Iout_{j}$ indicates the output current of the $j^{th}$ column. In figure~\ref{fig:resEq}, we refer vertical and horizontal slices of the crossbar array as Column Linear Systems (LSCols) and Row Linear Systems (LSRows), respectively. To demonstrate the formulation of $G_{non-ideal}$, we employ 6 major steps involving Equations 2 to 14. We next describe these steps in turn below.

{\bf\noindent Step 1: Formulate column linear systems}. We first formulate column linear systems ($LSCol_{1}$ to $LSCol_{N}$) using each vertical slice of the crossbar, shown in Figure~\ref{fig:resEq}. Let us consider the $j^{th}$ vertical slice corresponding to the $LSCol_{j}$ system (Equations 2 to 4). Using Kirchhoff's Current Law (KCL) at all nodes $b_{i,j}$ present in the $j^{th}$ column, we obtain Equations 2 and 3. Equations 2 and 3 are then combined to obtain the linear system in Equation 4. In the case of an MxN crossbar, we have N such linear systems ($LSCol_{1}$ to $LSCol_{N}$).


\begin{equation}
\centering
\begin{aligned}
   \begin{split}
 	& \textbf{A}_\textbf{j} * \textbf{Vcol}_\textbf{j} = \textbf{VAcol}_\textbf{j} - \textbf{J} * Iout_{j} 
	\end{split}
	\end{aligned}
	\label{eq:colAnalysis}
\end{equation}
\begin{equation}
\centering
\begin{aligned}
   \begin{split}
 	& Iout_{j}= \sum_{x=1}^{x=M} G_{x,j}V_{x,j}   
	\end{split}
	\end{aligned}
	\label{eq:OutputCurrent}
\end{equation}
\begin{equation}
\centering
\begin{aligned}
 	& (\textbf{A}_\textbf{j} + \textbf{J}*\textbf{K}_\textbf{j} ) \textbf{Vcol}_\textbf{j} = \textbf{VAcol}_\textbf{j}  
	\end{aligned}
	\label{eq:colSystem}
\end{equation}
where,
\vspace*{-4pt}
\begin{gather*}
\textbf{A}_\textbf{j}= \begin{bmatrix}
    -1 & G_{2,j}r_{col} & 2*G_{3,j}r_{col} & \dots & (M-1)*G_{M,j}r_{col} \\
    0 & -1 & G_{3,j}r_{col} & \dots & (M-2)*G_{M,j}r_{col} \\
    0 & 0 & -1 & \dots & (M-3)*G_{M,j}r_{col} \\
	\vdots & \vdots & \vdots & \ddots & \vdots \\
    0 & 0 & 0 & \dots &-1 \\
\end{bmatrix},
\\
\textbf{Vcol}_\textbf{j} = \begin{bmatrix}
    V_{1,j} \\
    V_{2,j} \\
    \vdots \\
    V_{M,j} \\	
\end{bmatrix},
 \textbf{VAcol}_\textbf{j} = \begin{bmatrix}
    Va_{1,j} \\
    Va_{2,j} \\
    \vdots \\
    Va_{M,j} \\	
\end{bmatrix}, 
 \textbf{J} = \begin{bmatrix}
    R_{sense}+(M-1)*r_{col} \\
    R_{sense}+(M-2)*r_{col}\\
    \vdots \\
    R_{sense}\\	
\end{bmatrix}
\\
\textbf{K}_\textbf{j}= \begin{bmatrix}
   G_{1,j} & G_{2,j} & \dots & G_{M,j}
\end{bmatrix}
\end{gather*}

\vspace*{+4pt}
{\bf\noindent Step 2: Merge column linear systems.} Next, the column linear systems ($LSCol_{1}$ to $LSCol_{N}$) are merged to form a larger Column Linear System (merged-LSCol) as shown in Equation 5 and 6. We achieve this by using the \emph{direct sum} ($\oplus$) matrix operation on matrices ($\textbf{A}_\textbf{j}$ + \textbf{J}*$\textbf{K}_\textbf{j}$) and $\textbf{K}_\textbf{j}$ to obtain block matrices \textbf{COLmat} and  \textbf{Gmat}, respectively. In Equation 5, \textbf{CVcol} and \textbf{CVAcol} are vectors formed by concatenating $\textbf{Vcol}_\textbf{j}$ and $\textbf{VAcol}_\textbf{j}$ vectors, respectively. Note that, the vectors $\textbf{Vcol}_\textbf{j}$ and $\textbf{VAcol}_\textbf{j}$ are obtained in Step 1 (Equations 2 and 4). Further, $\textbf{Iout}_\textbf{non-ideal}$ in Equation 6 is a vector representing the output currents.

\begin{equation}
\centering
\begin{aligned}
 	& \textbf{COLmat} * \textbf{CVcol} = \textbf{CVAcol}
	\end{aligned}
	\label{eq:CLS}
\end{equation}

\begin{equation}
\centering
\begin{aligned}
 	& \textbf{Gmat} * \textbf{CVcol} = (\textbf{Iout}_\textbf{non-ideal})^{T} 
	\end{aligned}
	\label{eq:GMAT_LS}
\end{equation}
where,
\vspace*{-4pt}
\begin{gather*}
\textbf{COLmat} = \underset{j \in 1,2,..,N}{\oplus} (\textbf{A}_\textbf{j} + \textbf{J}*\textbf{K}_\textbf{j} ) \\
\textbf{Gmat} = \underset{j \in 1,2,..,N}{\oplus} (\textbf{K}_\textbf{j})
\\
\textbf{Iout}_\textbf{non-ideal} = \begin{bmatrix} 
  Iout_{1}& Iout_{2}& \dots& Iout_{N}
\end{bmatrix}\\
\textbf{CVcol} = \begin{bmatrix} 
    \textbf{Vcol}_\textbf{1}^T& \textbf{Vcol}_\textbf{2}^T& \dots & \textbf{Vcol}_\textbf{N}^T\\
\end{bmatrix}^T 
\\
\textbf{CVAcol} = \begin{bmatrix} 
  \textbf{VAcol}_\textbf{1}^T& \textbf{VAcol}_\textbf{2}^T& \dots & \textbf{VAcol}_\textbf{N}^T\\
\end{bmatrix}^T
\end{gather*}

\vspace*{+4pt}
{\bf\noindent Step 3: Formulate row linear systems.} Similar to Step 1, the row linear systems ($LSrow_{1}$ to $LSrow_{M}$) are formulated considering horizontal slices of the crossbar. We use KCL at nodes $a_{i,j}$ present in the $i^{th}$ horizontal slice to obtain Equation 7 which represents the $LSrow_{j}$ system. In case of an MxN crossbar, we have M such row linear systems ($LSrow_{1}$ to $LSrow_{M}$).

\begin{equation}
\centering
\begin{aligned}
   \begin{split}
 	& \textbf{B}_\textbf{i} * \textbf{Vrow}_\textbf{i} = \textbf{VrowIN}_\textbf{i} - \textbf{VARow}_\textbf{i}  
	\end{split}
	\end{aligned}
	\label{eq:rowAnalysis}
\end{equation}
where,
\vspace*{-4pt}
\begin{gather*}
\textbf{B}_\textbf{i}= \begin{bmatrix}
    G_{i,1} & G_{i,2}  & G_{i,3} & \dots&  G_{i,N} \\
    G_{i,1} & G_{i,2}*2 & G_{i,3}*2 & \dots& G_{i,N}*2 \\
    G_{i,1} & G_{i,2}*2 & G_{i,3}*3 & \dots& G_{i,N}*3 \\
	\vdots & \vdots & \vdots & \ddots & \vdots \\
    G_{i,1} & G_{i,2}*2 & G_{i,3}*3 & \dots & G_{i,N}*N \\
\end{bmatrix}*r_{row},
\\
\textbf{Vrow}_\textbf{i} = \begin{bmatrix}
    V_{i,1} \\
    V_{i,2} \\
    \vdots \\
    V_{i,N} \\	
\end{bmatrix},
\textbf{VArow}_\textbf{i} = \begin{bmatrix}
    Va_{i,1} \\
    Va_{i,2} \\
     \vdots  \\
    Va_{i,N} \\	
\end{bmatrix},
\textbf{VrowIN}_\textbf{i} = \begin{bmatrix}
    Vin_{i} \\
    Vin_{i} \\
    \vdots \\
    Vin_{i} \\	
\end{bmatrix}
\end{gather*}

\vspace*{+4pt}
{\bf\noindent Step 4: Merge row linear systems.} Next, the row linear systems obtained in Step 3 are merged to obtain a larger Row Linear System (merged-LSrow) as shown in Equation 8. \textbf{ROWmat} is a block matrix obtained by performing the \emph{direct sum} ($\oplus$) matrix operation on the matrix ($\textbf{B}_\textbf{i}$). Moreover, \textbf{CVrowIN}, \textbf{CVrow}, and \textbf{CVArow} are vectors formed by concatenating vectors (obtained in Step 3) $\textbf{VrowIN}_\textbf{i}$, $\textbf{VArow}_\textbf{i}$, and $\textbf{Vrow}_\textbf{i}$, respectively. 	

\begin{equation}
\centering
\begin{aligned}
 	& \textbf{CVrowIN} - \textbf{ROWmat} * \textbf{CVrow} = \textbf{CVArow} 
	\end{aligned}
	\label{eq:RLS}
\end{equation}
where,
\vspace*{-4pt}
\begin{gather*}
\textbf{ROWmat} = \underset{i \in 1,2,..,M}{\oplus} (\textbf{B}_\textbf{i}) \\
\textbf{CVrow} = \begin{bmatrix} 
    \textbf{Vrow}_\textbf{1}^T& \textbf{Vrow}_\textbf{2}^T& \dots& \textbf{Vrow}_\textbf{N}^T \\
\end{bmatrix}^T 
\\
\textbf{CVArow} = \begin{bmatrix} 
    \textbf{VArow}_\textbf{1}^T& \textbf{VArow}_\textbf{2}^T& \dots& \textbf{VArow}_\textbf{N}^T \\
\end{bmatrix}^T
\\
\textbf{CVrowIN} = \begin{bmatrix} 
     \textbf{VrowIN}_\textbf{1}^T& \textbf{VrowIN}_\textbf{2}^T& \dots& \textbf{VrowIN}_\textbf{N}^T \\
\end{bmatrix}^T
\end{gather*}
				
\vspace*{+4pt}
{\bf\noindent Step 5: Eliminate internal variables.} Next, the vectors \textbf{CVAcol} and \textbf{CVcol} comprising of internal variables $Va_{i,j}$ and $V_{i,j}$, respectively, are eliminated. In order to eliminate these variables, we use the merged-LScol and merged-LSrow systems obtained in Step 2 and 4, respectively. However, the merged-LScol and merged-LSrow equations cannot be used directly due to the mismatch in their Right-Hand Sides (RHS) (\textbf{CVAcol} $\ne$ \textbf{CVArow}). We resolve this mismatch by performing elementary row operations on Equation 8 to obtain Equation 9. Note that, the \textbf{CVrowINA} vector and the \textbf{ROWmatA} matrix are obtained by performing row switching,~\emph{i.e.}, an elementary row operation, on the \textbf{CVrowIN} vector and the \textbf{ROWmat} matrix, respectively. Next, the \textbf{CVAcol} vector is eliminated using Equations 5 and 9 to obtain Equation 10. Subsequently, the \textbf{CVcol} vector is eliminated using Equations 6 and 10 to yield Equation 11. Note that, Equation 12 details the \textbf{NETmat} matrix introduced in Equation 11. 

\begin{equation}
\centering
\begin{aligned}
 	& \textbf{CVrowINA} - \textbf{ROWmatA} * \textbf{CVcol} = \textbf{CVAcol}  
	\end{aligned}
	\label{eq:RLS_rearranged}
\end{equation}
\begin{equation}
\centering
\begin{aligned}
 	& (\textbf{COLmat} +\textbf{ROWmatA}) * \textbf{CVcol} = \textbf{CVrowINA}  
	\end{aligned}
	\label{eq:CVAcol_eliminated}
\end{equation}
\begin{equation}
\centering
\begin{aligned}
 	& (\textbf{Iout}_\textbf{non-ideal})^{T} = \textbf{NETmat} * \textbf{CVrowINA}
	\end{aligned}
	\label{eq:NetmatEq}
\end{equation}
\begin{equation}
\centering
\begin{aligned}
&\textbf{NETmat} = \textbf{Gmat} * (\textbf{COLmat} +\textbf{ROWmatA})^{-1}
	\end{aligned}
	\label{eq:InverseEq}
\end{equation}

\vspace*{+4pt}
{\bf\noindent Step 6: Reduce matrix dimension.} Finally, we reduce the size of matrices \textbf{NETmat} and \textbf{CVrowINA} by leveraging a key property of the \textbf{CVrowINA} vector,~\emph{i.e.}, it contains repeated elements. Recall that, the \textbf{CVrowIN} vector is formed by concatenating the $\textbf{VrowIN}_\textbf{i}$ vectors (Step 4), and the \textbf{CVrowINA} vector is obtained by performing row switching operations on the \textbf{CVrowIN} vector. Since the $\textbf{VrowIN}_\textbf{i}$ vector (Step 3) has repeated elements, consequently, the vectors \textbf{CVrowIN} and \textbf{CVrowINA} also have repeated elements. Exploiting this property, the columns of the \textbf{NETmat} matrix that are to be multiplied by same elements in \textbf{CVrowINA} can be summed using elementary column operations to yield a compressed \textbf{NETmatC} matrix (shown in Equation 13). Moreover, removing redundancies in vector \textbf{CVrowINA} leads to the ${\textbf{Vin}_\textbf{non-ideal}}^{T}$ vector. Further, Equation 13 can be re-written as Equation 14 to obtain the $\textbf{G}_\textbf{non-ideal}$ matrix. Note that, $\textbf{G}_\textbf{non-ideal}$ is a function of ($\textbf{G}$, $R_{sense}$, $r_{col}$, and $r_{row}$), and therefore can be constructed using the intermediate matrices \textbf{COLmat}, \textbf{ROWmat}, and \textbf{Gmat}.

\begin{equation}
\centering
\begin{aligned}
 	& (\textbf{Iout}_\textbf{non-ideal})^{T} = \textbf{NETmatC} * 
	(\textbf{Vin}_\textbf{non-ideal})^T
	\end{aligned}
	\label{eq:NETmatCeq}
\end{equation}
\begin{equation}
\centering
\begin{aligned}
 	& \textbf{Iout}_\textbf{non-ideal} = \textbf{Vin}_\textbf{non-ideal} * \textbf{G}_\textbf{non-ideal}
	\end{aligned}
	\label{eq:GNIdealEq}
\end{equation}
where,
\vspace*{-4pt}
\begin{gather*}
\textbf{Vin}_\textbf{non-ideal} = \begin{bmatrix} 
  Vin_{1}& Vin_{2}& \dots & Vin_{M}\\
\end{bmatrix}\\
\textbf{G}_\textbf{non-ideal} =  \textbf{NETmatC}^T = f(\textbf{G},R_{sense},r_{row},R_{col})
\end{gather*}

%% file: sections/fcm_caffe.tex
{\bf \noindent} In this section, we present the overall RxNN framework that enables evaluation of large-scale DNNs on resistive crossbar systems. RxNN is a functional simulator obtained by modifying the Caffe~\cite{caffe} deep learning framework to mimic non-ideal vector-matrix multiplications realized on resistive crossbars. Caffe models the convolution and fully-connected layers of DNNs as matrix-matrix and vector-matrix multiplications. RxNN maps these matrix-matrix and vector-matrix multiplications to a resistive crossbar system and evaluates application-level accuracy of DNN inference operations. It takes the trained DNN network and weights, resistive crossbar system description and crossbar parameters as inputs, and evaluates the DNN inference operation using FCM models. RxNN's primary objective is to evaluate the application-level accuracy of DNNs, however, it is also capable of generating execution traces to enable performance and energy estimation. RxNN can also be used for model-in-the-loop re-training to improve DNN inference accuracy in the presence of non-idealities.

\begin{figure}[htb]
  \centering
  \vspace*{-6pt}
  \includegraphics[width=\columnwidth]{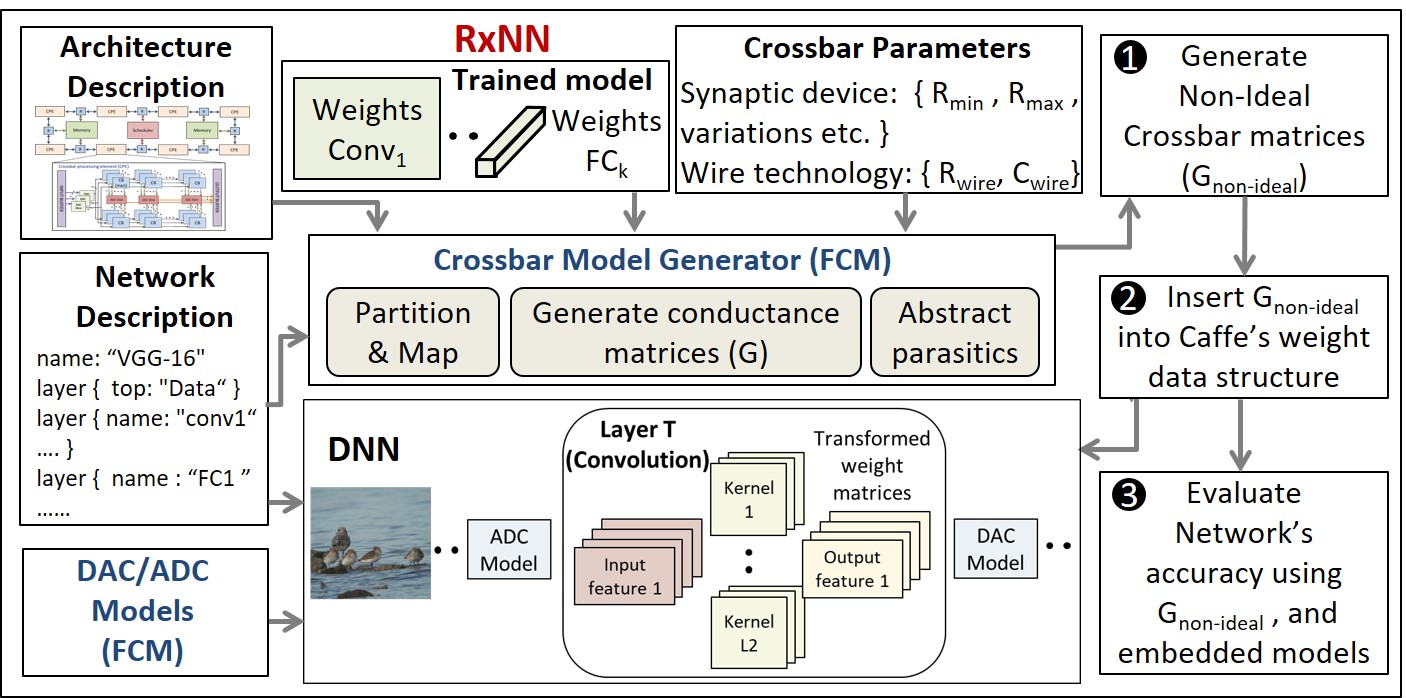}
  \vspace*{-10pt}
  \caption{ RxNN Overview}
  \label{fig:fcm-caffe}
  \vspace*{-5pt}
\end{figure}

Figure~\ref{fig:fcm-caffe} depicts the RxNN flow that consists of 3 steps. In step \circled{1}, RxNN maps the neural network to the specified target architecture. The weights are read from the trained Caffe model and virtually programmed into the crossbar array instances. Subsequently, the conductance matrices (G) corresponding to each resistive crossbar instance are generated, which are then transformed into the non-ideal conductance matrices ($G_{non-ideal}$) by abstracting crossbar non-idealities. Next, in step \circled{2}, the $G_{non-ideal}$ matrices associated with each DNN layer are incorporated back into the Caffe's original weight data structure. RxNN transparently utilizes Caffe's underlying data structures and optimized BLAS libraries, which is key to its performance and scalability. We note that steps \circled{1}-\circled{2} are performed only once for a given DNN and crossbar-based architecture. Thereafter, in step \circled{3}, RxNN evaluates the DNN for the given set of test inputs using embedded $G_{non-ideal}$ matrices and peripheral (ADC and DAC) models. During network evaluation, the DAC/ADC models are invoked as pre- and post-processing steps on the inputs/outputs of each convolutional and fully-connected layer. 

Next, we describe re-training with RxNN to improve the inference accuracy of DNNs on resistive crossbar systems. The major challenges that arise during DNN re-training for crossbar systems are: (i) the data-structures (inputs, outputs, weights) should abide by the range and resolution constraints at all times, and (ii) errors and gradients computed during back-propagation should be appropriately scaled to ensure network convergence~\footnote{The stochastic-gradient descent solver assumes the forward and backward passes to be contiguous and differentiable. However, crossbar abstraction of vector-matrix multiplication does not ensure these conditions.}. RxNN meets these constraints by utilizing a crossbar-based forward pass and a floating-point based backward pass. It appropriately converts and scales the data-structures between forward and backward passes to ensure that the network re-trains with minimal impact on the overall training time, which is extremely critical in the context of large-scale DNNs.

%% file: sections/exptsetup.tex
{\noindent} In this section, we describee the experimental setup used to evaluate the RxNN framework.

\begin{figure}[htb]
  \centering
  \vspace*{-8pt}
  \includegraphics[width=\columnwidth]{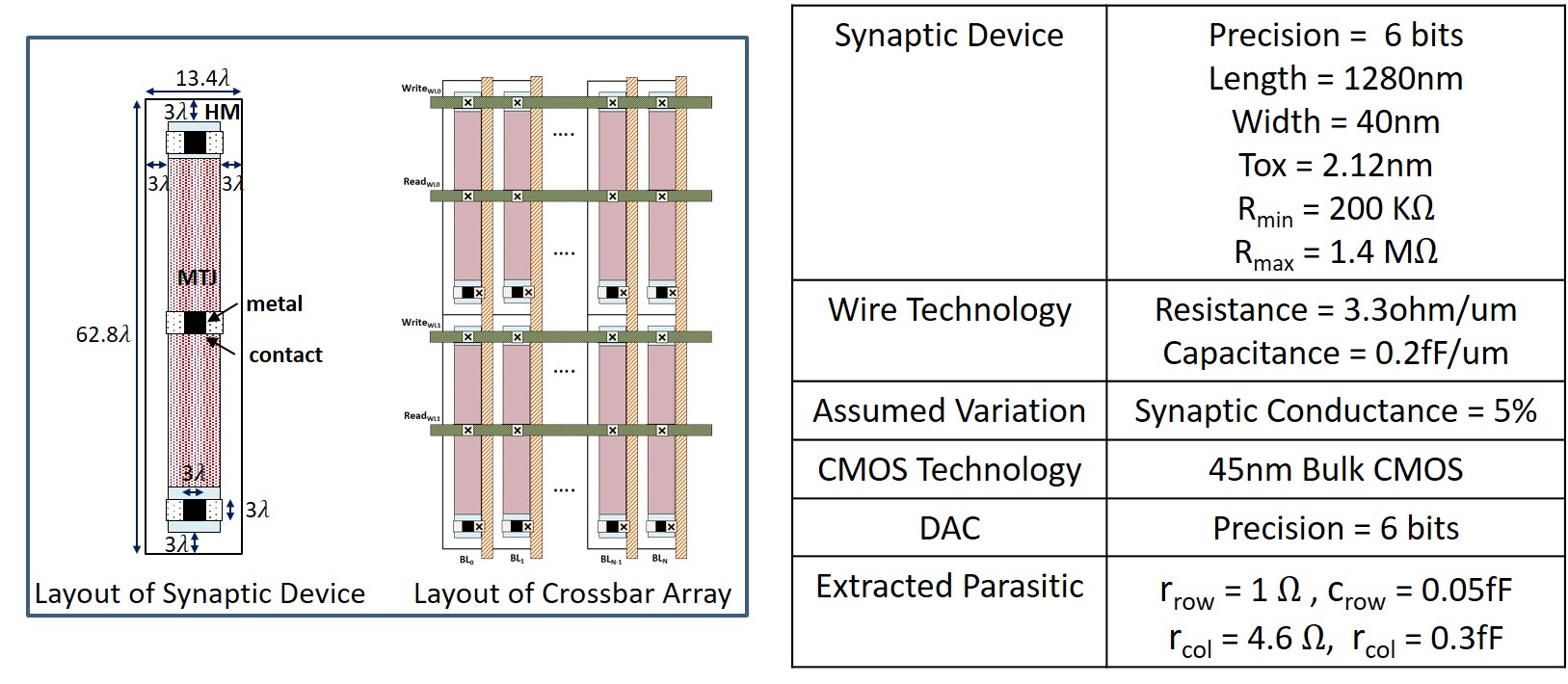}
  \vspace*{-16pt}
  \caption{ Device and Technology Parameters}
  \label{fig:crossbar_layout}
  \vspace*{-8pt}
\end{figure}

{\bf \noindent Device/Circuit simulation.} We use an in-house device model of the synaptic element~\cite{sengupta_synapseMain} that is based on the solution of Landau-Lifshitz-Gilbert (LLG) magnetization dynamics and Non-Equilibrium-Green’s Function (NEGF) electron transport. Circuit-level simulations are performed in HSPICE using the 45nm bulk CMOS technology and the synaptic device model. Our simulations use the ADC and DAC circuits proposed in~\cite{adc_ibm,verma}. The interconnect parasitics ($r_{row}$, $r_{col}$) are extracted using the device and crossbar array layouts. Figure~\ref{fig:crossbar_layout} shows these layouts that are performed using the design rules specified in~\cite{rulesmosis}. The table in Figure~\ref{fig:crossbar_layout} details the device, technology~\cite{45nm_params_intel}, and variation parameters~\cite{synapse_variation} assumed in our experiments. We also characterize a resistive crossbar array to compute energy at the crossbar-level which is used as a technology parameter in RxNN to estimate system-level energy consumption.


\begin{table}[htb]
  \centering
  \vspace*{-2pt}
  \caption{Benchmark DNN Applications}
  \includegraphics[width=\columnwidth]{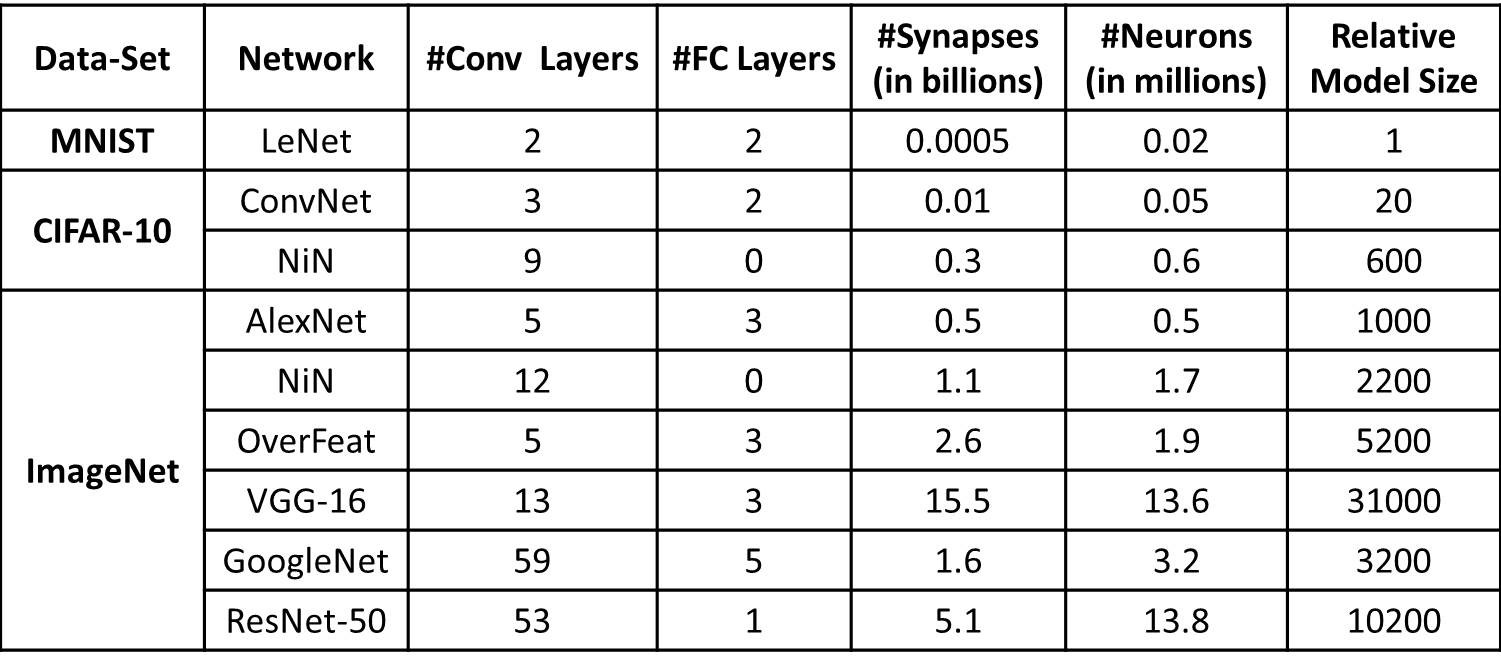}
  \vspace*{-6pt}
  \label{tab:benchmark}
  \vspace*{-6pt}
\end{table}

{\bf \noindent  Application-Level simulation.} We evaluated the application-level accuracy and energy of several popular DNNs on the resistive crossbar system using RxNN. Table~\ref{tab:benchmark} provides details of the benchmark DNNs, including the number of convolution and fully-connected layers, the targeted data-set, and the number of neurons and synaptic connections. We also present the relative model size to highlight the difference between these benchmark DNNs. To evaluate energy consumption, we use an architecture similar to~\cite{spindle}. 


%% file: sections/results.tex
\noindent We now present the experimental results to demonstrate the modeling accuracy and speedups achieved by FCM over circuit simulation. We also evaluate the application-level accuracy of large-scale DNNs on non-ideal resistive crossbar systems using RxNN.

\begin{figure}[htb]
  \centering
  \vspace*{-5pt}
  \includegraphics[width=\columnwidth]{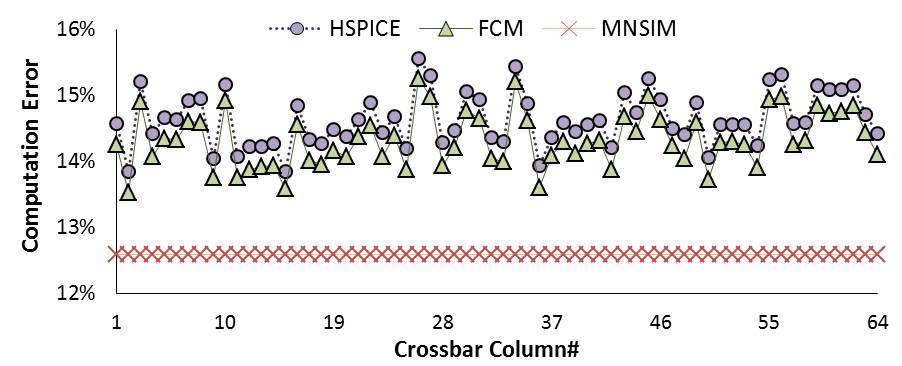}
  \vspace*{-10pt}
  \caption{ Computation Errors observed in crossbar for various crossbar models  }
  \label{fig:fcm_accuracy}
  \vspace*{-16pt}
\end{figure}

\subsection{FCM: Crossbar-level Evaluation}
\label{subsec:fcmresult}

{\bf \noindent Modeling Accuracy.} Figure~\ref{fig:fcm_accuracy} shows the errors in vector-matrix multiplications realized using a 64x64 non-ideal crossbar. We compute errors using three different crossbar models, \emph{viz.}, HSPICE, FCM, and MNSIM~\cite{mnsim}. The X-axis represents the crossbar column, and the Y-axis depicts the error incurred due to non-idealities in the vector-matrix multiplication. We observe that the simple error model (MNSIM) deviates considerably from the HSPICE model. This is expected, as it does not consider the dependence of errors on several factors including the applied inputs, the crossbar state, and the crossbar column. In contrast, the FCM model considers these dynamic factors and is therefore able to closely match the HSPICE model. The maximum deviation between the errors estimated by MNSIM and the  errors computed using HSPICE is about 3.51\%. In the case of FCM, the maximum deviation is found to be 0.28\%, which is significantly smaller. 


\vspace*{+4pt}

\begin{figure}[htb]
  \vspace*{-0pt}
  \centering
  \includegraphics[width=0.5\columnwidth]{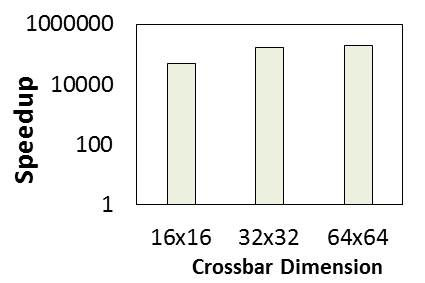}
  \vspace*{-4pt}
  \caption{FCM speedup over HPSICE}
  \label{fig:speedup}
  \vspace*{-10pt}
\end{figure}

{\bf \noindent Speedup.} To evaluate the speedup of FCM over HSPICE, we measure the execution time of FCM and HSPICE  for various crossbar sizes. Figure~\ref{fig:speedup} details the speedup achieved using FCM over HSPICE. We observe a speedup of about 5 orders in magnitude. Moreover, as expected, the speedup increases for larger crossbar arrays.

\vspace*{+4pt}
{\bf \noindent Model generation overhead.} Recall that FCM's crossbar model generator transforms the weight matrix (W) to a non-ideal conductance matrix ($G_{non-ideal}$), which incurs a one-time overhead. In our evaluation, we found the modeling overhead to be 0.038, 1.2, and 61 seconds for 16x16, 32x32, and 64x64 crossbar arrays, respectively. While considerable for larger crossbars, these one-time overheads are amortized over a large number of inference operations typically processed by RxNN. 

\subsection{RxNN: Application-Level Evaluation }
\label{subsec:fcm-cafferesult}

\begin{figure}[htb]
  \centering
  \vspace*{-5pt}
  \includegraphics[width=\columnwidth]{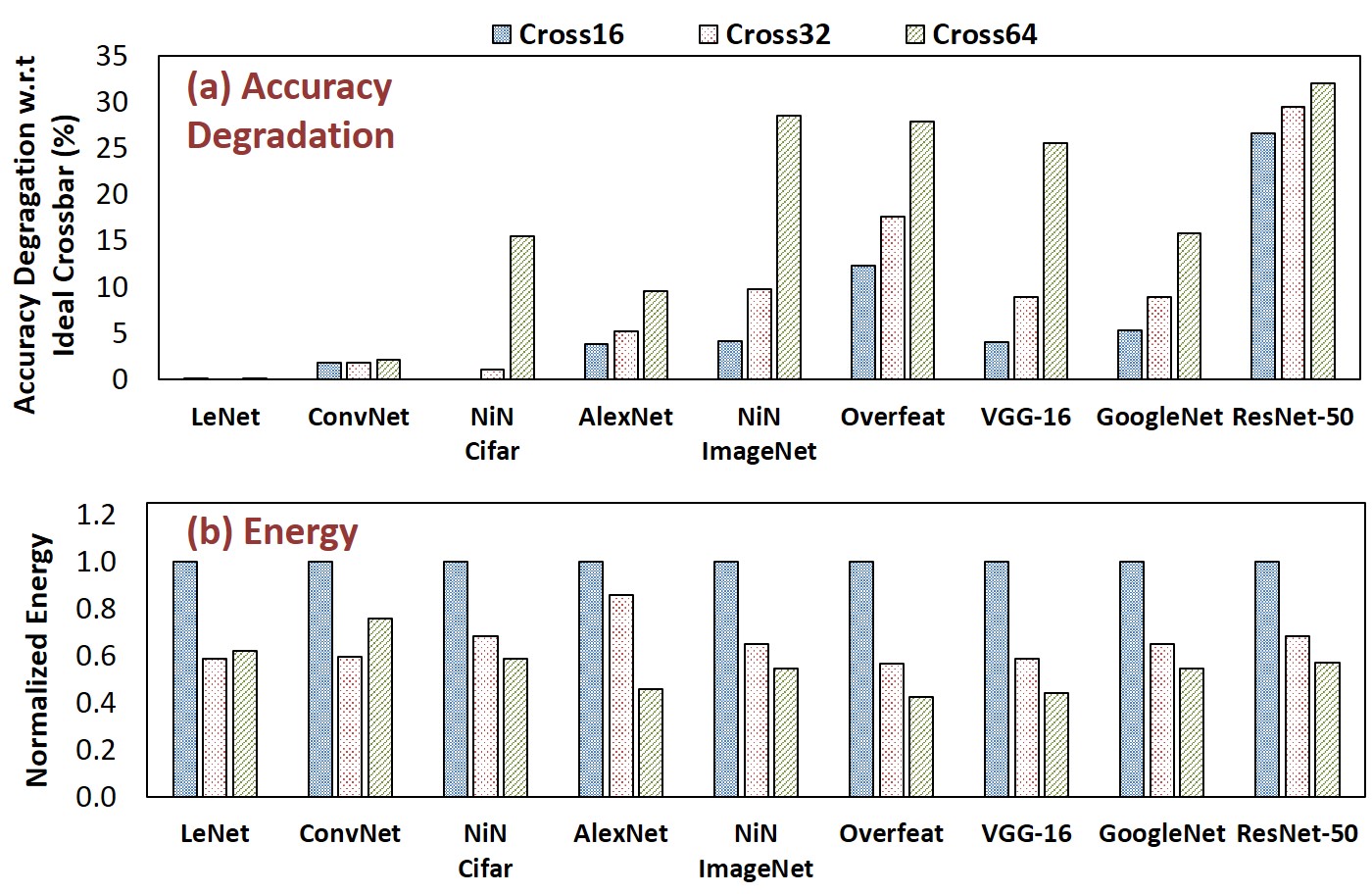}
  \vspace*{-10pt}
  \caption{Application-Level evaluation using RxNN}
  \label{fig:fcm-caffe-result}
  \vspace*{-5pt}
\end{figure}

Next, we apply RxNN to evaluate the accuracy degradation due to crossbar non-idealities at the application-level for the benchmark DNNs. We implement three different resistive crossbar systems designed using crossbars of size 16x16 (Cross16), 32x32 (Cross32), and 64x64 (Cross64). Figure~\ref{fig:fcm-caffe-result}(a) shows the accuracy degradation for these designs with respect to our baseline, {\em i.e.}, an ideal crossbar with no device and circuit-level non-idealities. We first compare the accuracy degradation of the Cross64 design across DNNs. We observe that for simple networks (LeNet and ConvNet) the accuracy degradation due to non-idealities is quite small. For example, LeNet and ConvNet networks suffer accuracy degradation of 0.05\% and 2.2\%, respectively. In contrast,  the accuracy loss due to non-idealities is considerable for large-scale DNNs. For instance, VGG-16, OverFeat, and Resnet-50 networks incur accuracy losses of 25.6\%, 27.8\%, and 32\%, respectively. We observe similar accuracy degradation trend across simple and large-scale DNNs for the Cross16 and Cross32 designs as well.

Next, we compare the accuracy degradation across designs with different crossbar sizes (Cross16, Cross32, and Cross64). As evident from Figure~\ref{fig:fcm-caffe-result}(a), the accuracy degradation for the Cross16 design is less than the Cross32 design, which is in turn less than the Cross64 design. This trend is expected as the impact of non-idealities is lower for smaller crossbar arrays (Section~\ref{subsec:NI_evaluation}). However, smaller crossbar arrays are not desirable in terms of energy efficiency. Figure~\ref{fig:fcm-caffe-result}(b) depicts the normalized energy consumption per image for the Cross16, Cross32, and Cross64 designs. The Cross16 design consumes higher energy than the Cross32 design, which in turn consumes higher energy than the Cross64 design for most cases. Since the major components of the energy consumed in resistive crossbar systems are peripherals (ADCs and DACs),larger crossbar arrays that amortize the energy cost of ADCs and DACs over more columns and rows have superior energy efficiency. Note that, for the LeNet and ConvNet DNNs, the energy of the Cross64 design is higher than the Cross32 design. This is because the crossbars in the Cross64 design are under-utilized in case of these relatively small networks. Therefore, Cross64 suffers from energy overheads due to redundant computations performed in the unused rows/columns.  

\begin{figure}[htb]
  \centering
  \vspace*{-6pt}
  \includegraphics[width=0.7\columnwidth]{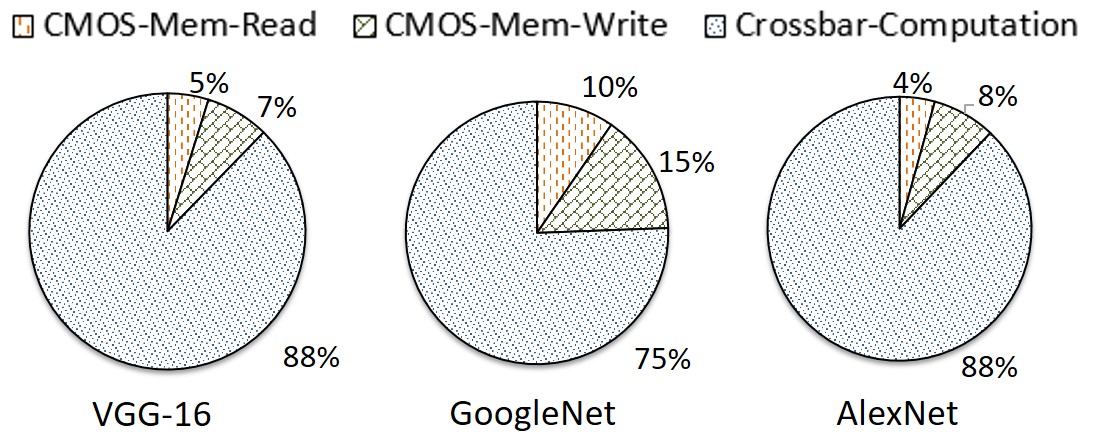}
  \vspace*{-2pt}
  \caption{ Energy Breakdown for Cross64 implementation}
  \label{fig:energyBreakdown}
  \vspace*{-6pt}
\end{figure}

We next present the energy breakdown of three networks,~\emph{viz.}, VGG-16, GoogleNet, and AlexNet realized on the Cross64 design. Figure~\ref{fig:energyBreakdown} shows the energy breakdown of these networks considering -- read energy for inputs (CMOS-Mem-Read), write energy for outputs (CMOS-Mem-Write), and computation energy for vector-matrix multiplications (Cross-Computation). We observe that the major energy component is the vector-matrix multiplications (Cross-Computation) which is in turn dominated by the ADCs and DACs.

In summary, there exists a fundamental trade-off between the application-level accuracy and the system energy which needs to be examined, in order to determine the architectures for future resistive crossbar systems. RxNN intends to drive these decisions by providing a software platform that can precisely evaluate crossbar architectures executing large-scale DNNs.

{\bf \noindent RxNN speed {\em vs.} Caffe.}
We also evaluated the slowdown of RxNN with respect to the baseline Caffe framework (without any crossbar modeling), and found that it amounts to 2.5X and 2.75X for inference and re-training, respectively, across our benchmark applications. We believe this is a reasonable overhead given the highly optimized nature of Caffe, and the fact that much like Caffe, RxNN can also leverage multi-cores, GPUs, and clusters for increased processing throughput.

\subsection{Sensitivity of accuracy to non-idealities}
\label{subsec:sensitivityAnalysis}

\begin{figure}[htb]
  \centering
  \vspace*{-6pt}
  \includegraphics[width=0.95\columnwidth]{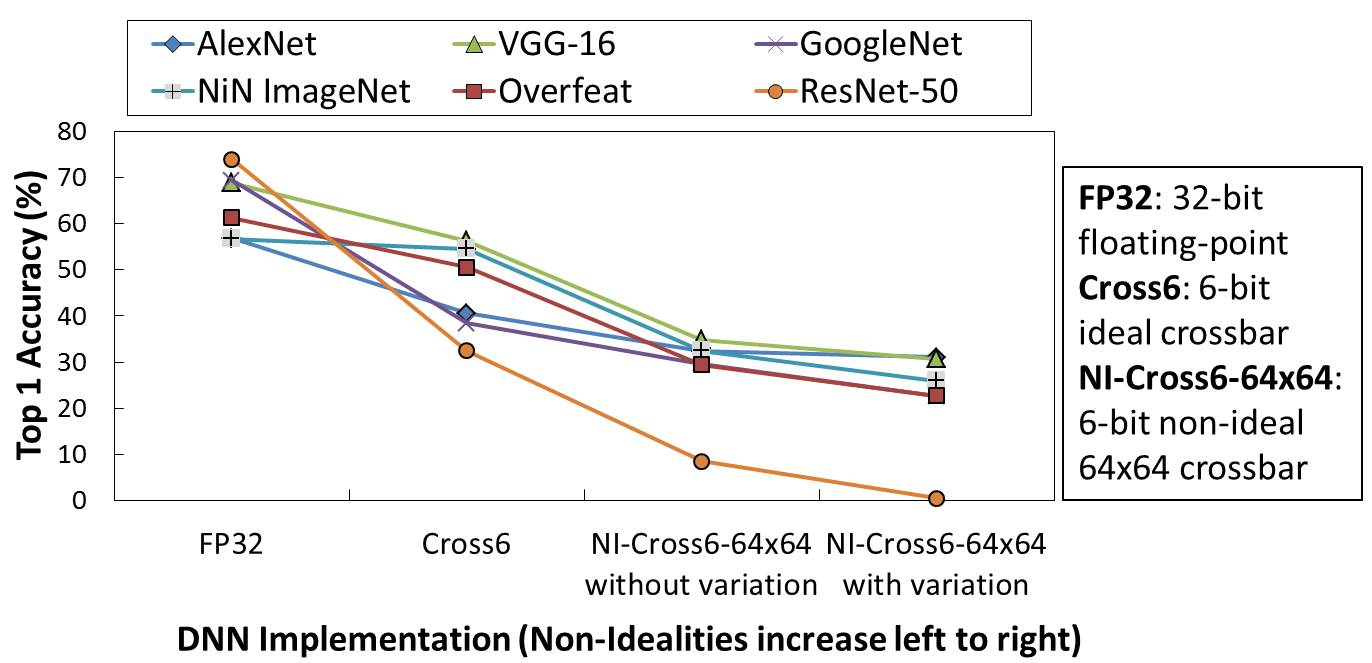}
  \vspace*{-4pt}
  \caption{ Accuracy's sensitivity to non-idealities}
  \label{fig:accurracySensitivity}
  \vspace*{-20pt}
\end{figure}

To further illustrate the impact of non-idealities on the application-level accuracy, we present a sensitivity analysis in Figure~\ref{fig:accurracySensitivity}. We plot the accuracies of 6 large-scale networks,~\emph{viz.}, AlexNet, VGG-16, GoogleNet, NiN, Overfeat, and ResNet-50 for implementations differing in their degree of non-idealities. The implementations that we use are: (i) floating-point implementation realized on an x86 CPU architecture (FP32), (ii) 6-bit ideal crossbar design (Cross6) without any crossbar non-idealities, and (iii) 6-bit non-ideal crossbar based designs with and without variations (NI-Cross6-64x64). Note that the FP32 CPU-based software implementation does not use crossbars and hence does not suffer from any non-idealities. As shown in Figure~\ref{fig:accurracySensitivity}, the accuracy drops from left to right as more non-idealities are incorporated. We observe two significant accuracy drops, one between FP32 and Cross6 implementations, and other between Cross6 and NI-Cross6-64x64 implementations. The degradation between FP32 and Cross6 is due to the limited precision of the synaptic devices, ADCs, and DACs. 
In contrast, the drop in accuracy from Cross6 to NI-Cross6-64x64 is due to the device and circuit-level non-idealities.  

\subsection{Re-training DNNs using RxNN}
\label{subsec:retrainingResult}

\begin{figure}[htb]
  \centering
  \vspace*{-10pt}
  \includegraphics[width=0.70\columnwidth]{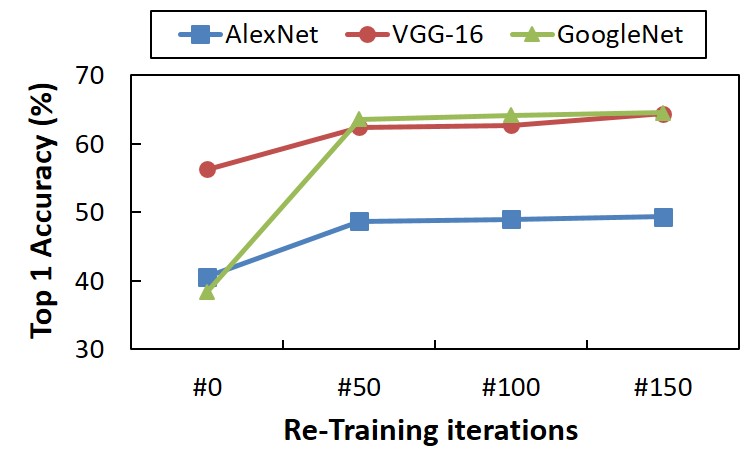}
  \vspace*{-2pt}
  \caption{ Re-training using RxNN}
  \label{fig:retrainingResult}
  \vspace*{-20pt}
\end{figure}

Next, we show the effectiveness of RxNN in re-training large-scale DNNs for resistive crossbar systems. To that end, we re-trained three networks,~\emph{viz.}, AlexNet, VGG-16, and GoogleNet, as shown in Figure~\ref{fig:retrainingResult}. Our experiments show that with only 150 iterations of re-training RxNN can achieve $\sim$9\%, $\sim$8\%, $\sim$26\%  improvement in accuracy for AlexNet, VGG-16, and GoogleNet, respectively. Notwithstanding these improvements, there is still a substantial drop in accuracy that cannot be recovered by re-training alone, calling for additional error mitigation and compensation techniques.


\subsection{Visualizing the impact of non-idealities}
\label{subsec:sensitivityAnalysis}

\begin{figure}[htb] 
 \centering
   \vspace*{-6pt} 
   \includegraphics[width=0.70\columnwidth]{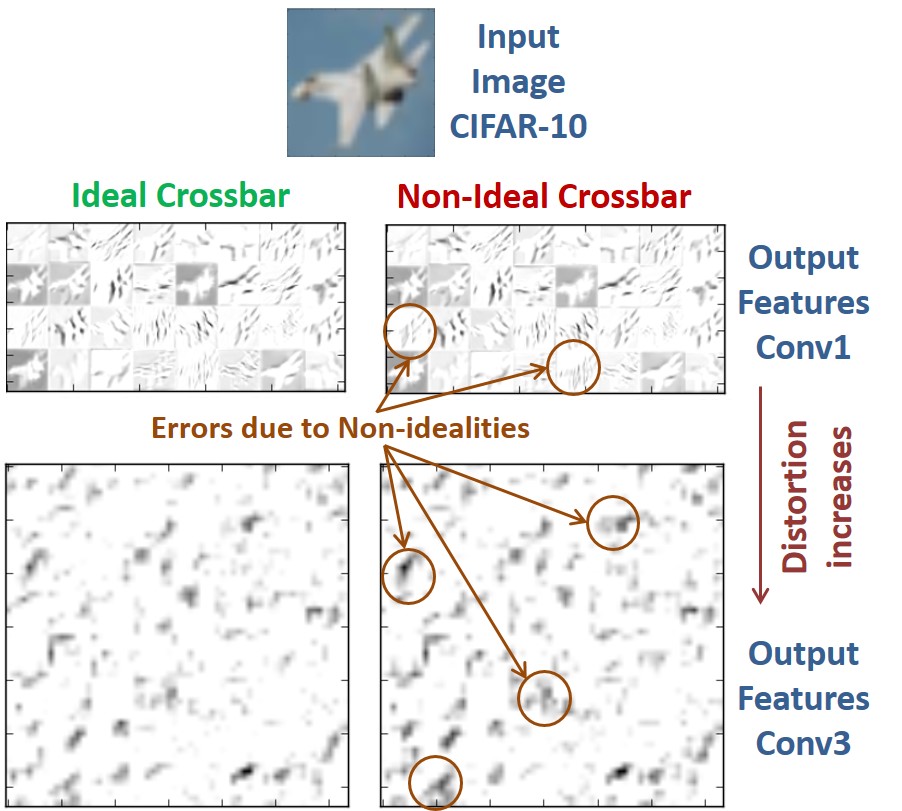}
  \vspace*{-0pt} 
 \caption{Visual demonstration of errors using ConvNet}
  \label{fig:DNNError}
  \vspace*{-10pt}
\end{figure}

To provide further insights into the impact of non-idealities at the application-level, Figure~\ref{fig:DNNError} compares the output features obtained from an ideal resistive crossbar system and a non-ideal resistive crossbar system for two convolution layers (Conv1 and Conv3) of the ConvNet network executing on the CIFAR-10 dataset. Some of the significant distortions in features are highlighted in the figure using circles. We observe that the impact of non-idealities increases considerably as we go deeper into the network (Conv3 layer outputs show increased artefacts compared to Conv1 layer outputs in Figure~\ref{fig:DNNError}). This is consistent with the observation from Figure~\ref{fig:fcm-caffe-result}(a) that deeper DNNs show greater degradation in accuracy due to crossbar non-idealities.

In summary, our results underscore the utility of RxNN in evaluating and re-training large-scale DNNs on resistive crossbar architectures. They also motivate the need for further research into techniques to mitigate and compensate the effects of crossbar non-idealities in the context of large-scale DNNs.



%% file: sections/conclusion.tex
{\bf \noindent} Resistive crossbars realized using non-volatile meemory devices promise to enable compact, energy-efficient hardware for DNNs. In this work, we evaluate the impact of various device and circuit non-idealities that are present in crossbars on the overall accuracy of large-scale DNNs. We propose FCM, a fast and accurate model to evaluate vector-matrix multiplications realized on resistive crossbars. Using FCM, we construct RxNN, a software simulation framework to evaluate large-scale DNNs on resistive crossbar systems. Our experiments with RxNN indicate that accuracy degradation due to non-idealities is a significant concern for large-scale DNNs. Re-training can only partly restore the accuracy lost, necessitating a need for further error mitigation and compensation schemes.

%% file: biographies.tex
\vspace*{-30pt}
\begin{IEEEbiography}
[{\includegraphics[width=1.1in,height=1.25in,clip,keepaspectratio]{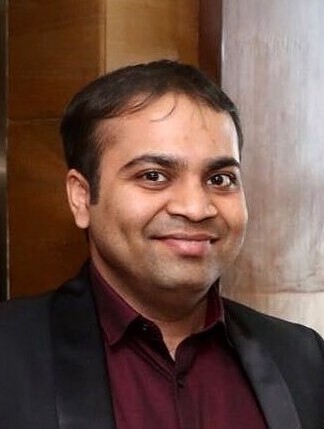}}]
{Shubham Jain} is currently a research staff member at IBM T.J. Watson Research Center, Yorktown Heights, New York. He has a B.Tech (Hons.) degree in Electronics and Electrical Communication Engineering from the Indian Institute of Technology, Kharagpur, India, in 2012, and a Ph.D. degree in Electrical and Computer Engineering from Purdue University, West Lafayette, Indiana, in 2019. His primary research interests include AI hardware, architecture for post-CMOS devices, in-memory computing, and approximate computing. Previously, he worked as a design engineer in the Bangalore Design Center, Qualcomm, Bangalore, India from 2012 to 2014. He also worked as a summer intern at IBM T.J Watson Research Center, Yorktown Heights, in 2017 and 2018. He has received the Mitacs Globalink scholarship from Mitacs, in 2011, the Andrews Fellowship from Purdue University, in 2014, and the A. Richard Newton Young Student Fellowship from DAC in 2015. His research has received the best technical paper award in DAC 2018, and a best-in session award in TECHCON 2016.
\end{IEEEbiography}
\vspace*{-30pt} 
\begin{IEEEbiography}
[{\includegraphics[width=1.1in,height=1.25in,clip,keepaspectratio]{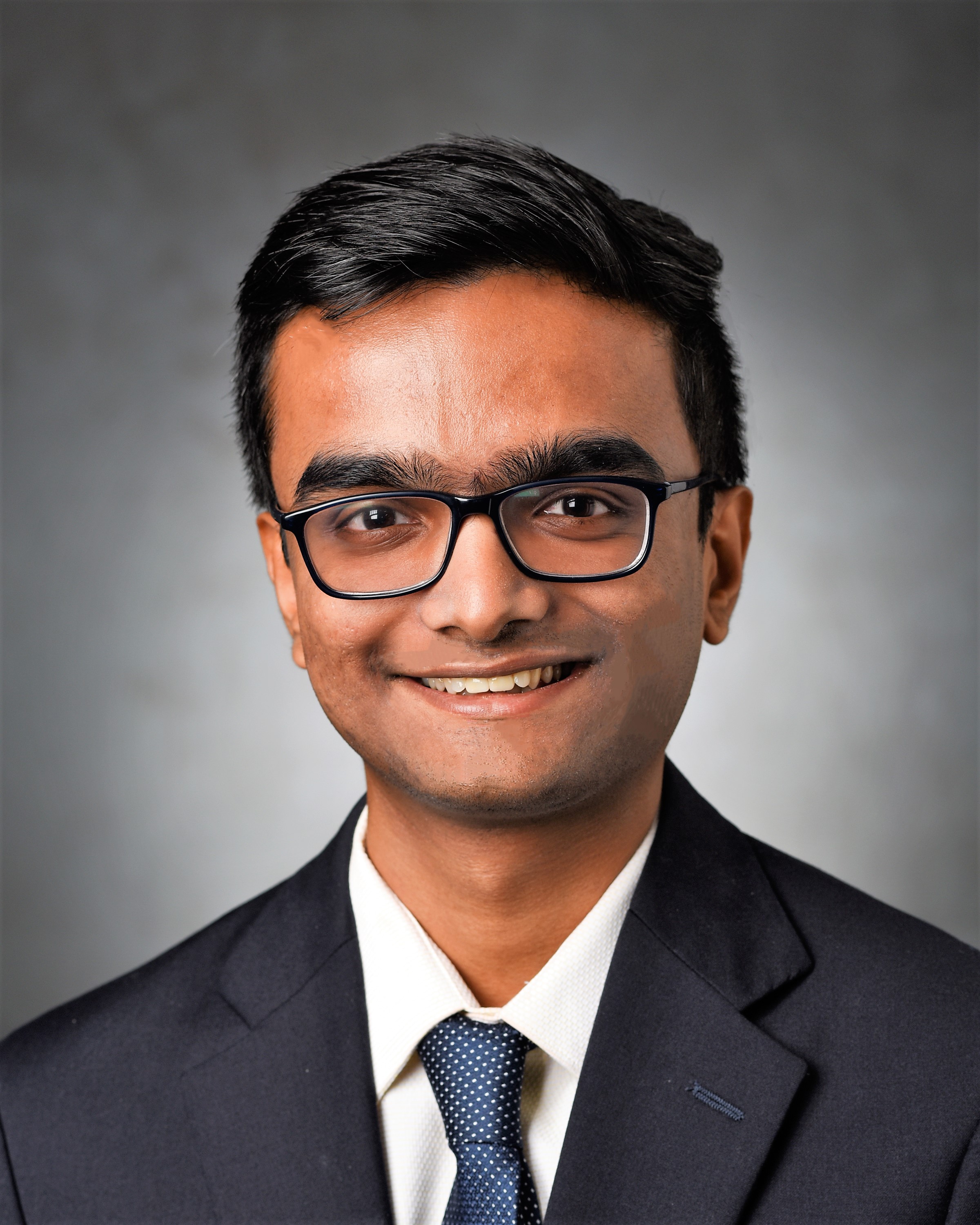}}]
{Abhronil Sengupta} is an Assistant Professor in the School of Electrical Engineering and Computer Science at Penn State University. He received the PhD degree in Electrical and Computer Engineering from Purdue University in 2018 and the B.E. degree from Jadavpur University, India in 2013. He worked as a DAAD (German Academic Exchange Service) Fellow at the University of Hamburg, Germany in 2012, and as a graduate research intern at Circuit Research Labs, Intel Labs in 2016 and Facebook Reality Labs in 2017.

Prof. Sengupta is pursuing an inter-disciplinary research agenda at the intersection of hardware and software across the stack of sensors, devices, circuits, systems and algorithms for enabling low-power event-driven cognitive intelligence. He has published over 45 articles in referred journals and conferences and holds 4 granted/pending US patents. He serves on the Technical Program Committee of Design Automation Conference (DAC 2019), International Symposium on Quality Electronic Design (ISQED 2019) and ACM Great Lakes Symposium on VLSI (GLSVLSI 2019). He has been awarded the IEEE SiPS Best Paper Award (2018), Schmidt Science Fellows Award nominee (2017), Bilsland Dissertation Fellowship (2017), CSPIN Student Presenter Award (2015), Birck Fellowship (2013) and the DAAD WISE Fellowship (2012).
\end{IEEEbiography}
\vspace*{-25pt}
\begin{IEEEbiography}
[{\includegraphics[width=1.1in,height=1.25in,clip,keepaspectratio]{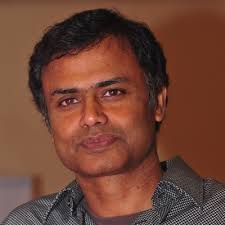}}]
{Kaushik Roy} received the BTech degree in electronics
and electrical communications engineering
from the Indian Institute of Technology, Kharagpur,
India, and the PhD degree from the Department
of Electrical and Computer Engineering,
University of Illinois at Urbana-Champaign in
1990. He was with the Semiconductor Process
and Design Center of Texas Instruments, Dallas,
where he worked on FPGA architecture development
and low-power circuit design. He joined the
electrical and computer engineering faculty at Purdue
University, West Lafayette, IN, in 1993, where he is currently Edward
G. Tiedemann Jr. Distinguished Professor. His research interests include
spintronics, device-circuit co-design for nano-scale Silicon and non-Silicon
technologies, low-power electronics for portable computing and wireless
communications, and new computing models enabled by emerging
technologies. He has published more than 600 papers in refereed journals
and conferences, holds 15 patents, graduated 60 PhD students, and
is coauthor of two books on Low Power CMOS VLSI Design (Wiley \&
McGraw Hill). He received the US National Science Foundation Career
Development Award in 1995, IBM faculty partnership award, ATT/Lucent
Foundation award, 2005 SRC Technical Excellence Award, SRC Inventors
Award, Purdue College of Engineering Research Excellence Award,
Humboldt Research Award in 2010, 2010 IEEE Circuits and Systems
Society Technical Achievement Award, Distinguished Alumnus Award
from Indian Institute of Technology, Kharagpur, Fulbright-Nehru Distinguished
Chair, and Best Paper Awards at 1997 International Test Conference,
IEEE 2000 International Symposium on Quality of IC Design, 2003
IEEE Latin American Test Workshop, 2003 IEEE Nano, 2004 IEEE International
Conference on Computer Design, 2006 IEEE/ACM International
Symposium on Low Power Electronics \& Design, and 2005 IEEE Circuits
and System Society Outstanding Young Author Award (Chris Kim), 2006
IEEE Transactions on VLSI Systems Best Paper Award, 2012 ACM/IEEE
International Symposium on Low Power Electronics and Design Best
Paper Award, 2013 IEEE Transactions on VLSI Best Paper Award. He
was a Purdue University Faculty scholar (1998-2003). He was a
Research Visionary board member of Motorola Labs (2002) and held the
M.K. Gandhi Distinguished Visiting faculty at Indian Institute of Technology
(Bombay). He has been in the editorial board of IEEE Design and
Test, IEEE Transactions on Circuits and Systems, IEEE Transactions on
VLSI Systems, and IEEE Transactions on Electron Devices. He was the
guest editor for Special Issue on Low-Power VLSI in the IEEE Design and
Test (1994) and IEEE Transactions on VLSI Systems (June 2000), IEE
Proceedings—Computers and Digital Techniques (July 2002), and IEEE
Journal on Emerging and Selected Topics in Circuits and Systems
(2011). He is a fellow of the IEEE
\end{IEEEbiography}

\vspace*{-25pt}
\begin{IEEEbiography}
[{\includegraphics[width=1.1in,height=1.25in,clip,keepaspectratio]{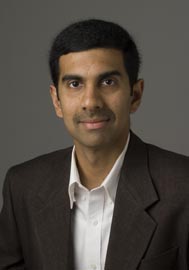}}]
{Anand Raghunathan} 
is a Professor of Electrical and Computer Engineering and Chair of the VLSI area at Purdue University, where he directs research in the Integrated Systems Laboratory. His current areas of research include domain-specific architecture, system-on-chip design, computing with post-CMOS devices, and heterogeneous parallel computing. Previously, he was a Senior Research Staff Member at NEC Laboratories America, where he led projects on system-on-chip architecture and design methodology. He has also held the Gopalakrishnan Visiting Chair in the Department of Computer Science and Engineering at the Indian Institute of Technology, Madras. 

Prof. Raghunathan has co-authored a book, eight book chapters, and over 200 refereed journal and conference papers, and holds 21 U.S patents. His publications received eight best paper awards and five best paper nominations. He received a Patent of the Year Award and two Technology Commercialization Awards from NEC, and was chosen among the MIT TR35 (top 35 innovators under 35 years across various disciplines of science and technology) in 2006.

Prof. Raghunathan has been a member of the technical program and organizing committees of several leading conferences and workshops, chaired premier IEEE/ACM conferences (CASES, ISLPED, VTS, and VLSI Design), and served on the editorial boards of various IEEE and ACM journals in his areas of interest. He received the IEEE Meritorious Service Award and Outstanding Service Award. He is a Fellow of the IEEE and Golden Core Member of the IEEE Computer Society. Prof. Raghunathan received the B. Tech. degree from the Indian Institute of Technology, Madras, and the M.A. and Ph.D. degrees from Princeton University.
\end{IEEEbiography}